\documentclass[aps,preprint,amsmath,noshowkeyws,noshowpacs,superscriptaddress,prb]{revtex4-2}
\usepackage{amssymb}
\usepackage{amsfonts}
\usepackage[verbose]{placeins}
\usepackage{graphicx}% Include figure files
\usepackage{dcolumn}% Align table columns on decimal point
\usepackage{bm}% bold math
\usepackage{multirow}
\usepackage[version=3]{mhchem}
\usepackage{tikz}
\usepackage{setspace} % set spacing
\usepackage{acronym} % used for adding accronyms
\usepackage[acronym]{glossaries}
\makeglossaries
\printglossary[type=\acronymtype]
\acrodef{BSE}[BSE]{Bethe-Salpeter equation}
\acrodef{C2DB}[C2DB]{computational 2D materials database}
\acrodef{CBE}[CBE]{conduction band edge}
\acrodef{DFT}[DFT]{density functional theory}
\acrodef{PBE}[PBE]{Perdew-Burke-Ernzerhof}
\acrodef{VBE}[VBE]{valence band edge}
\acrodef{TMDC}[TMDC]{transition metal dichalcogenide}
\acrodef{BZ}[BZ]{Brillouin zone}
\acrodef{SM}[SM]{supporting material}
\acrodef{SOC}[SOC]{spin-orbit coupling}
\acrodef{HSE06}[HSE06]{Heyd-Scuseria-Ernzerhof}
\acrodef{2DM}[2DM]{two-dimensional monolayer material}
\acrodef{YSH}[YSH]{Yukawa screened hybrid}
\acrodef{IPR}[IPR]{inverse participation ratio}
\acrodef{TM}[TM]{transition metal}
\acrodef{BEC}[BEC]{Born effective charge}
\acrodef{VASP}[VASP]{Vienna \textit{ab initio} simulation package}

\usepackage[utf8]{inputenc}
\usepackage{longtable} %Multi-pages table
\usepackage{float}
\usepackage[linktocpage=true]{hyperref}% add hypertext capabilities
\usepackage[bottom]{footmisc} 
\usepackage{amsmath,esint}
\usepackage{verbatim}
\usepackage{spverbatim}

\makeatletter
\def\maketitle{
\@author@finish
\title@column\titleblock@produce
\suppressfloats[t]}
\makeatother
\RequirePackage[normalem]{ulem}
\RequirePackage{color}
\definecolor{RED}{rgb}{1,0,0}
\definecolor{BLUE}{rgb}{0,0,1}
\definecolor{orange}{rgb}{1,0.65,0}
\definecolor{green}{rgb}{0,0.6,0}

\usepackage{comment}
\usepackage{soul}
\usepackage{notes2bib}
\usepackage{hyperref}
\hypersetup{
  pdfnewwindow=true, 
  colorlinks=true,
  linkcolor=blue, 
  anchorcolor=blue,
  citecolor=blue, 
  filecolor=blue,
  menucolor=blue, 
  urlcolor=blue}
\usepackage{orcidlink}
\usepackage{bbm} % lower case double stroke

\begin{document}

%\preprint{APS/123-QED}
\title{Efficiency of band edge optical transitions of 2D monolayer materials: A high-throughput computational study}% Force line breaks with 

\author{A. F. G{\'o}mez-Bastidas\orcidlink{0009-0004-3689-9449}}
\email{gomezbaa@mcmaster.ca}
\affiliation{Department of Materials Science and Engineering, \href{https://ror.org/02fa3aq29}{McMaster University}, 1280 Main Street West, Hamilton, Ontario L8S 4L8, Canada}

\author{Karthik Sriram\orcidlink{0009-0000-0701-1878}}
\affiliation{Department of Materials Science and Engineering, \href{https://ror.org/02fa3aq29}{McMaster University}, 1280 Main Street West, Hamilton, Ontario L8S 4L8, Canada}
\affiliation{Department of Metallurgical and Materials Engineering, \href{https://ror.org/03v0r5n49}{Indian Institute of Technology Madras}, Chennai 600036, India}

\author{A. C. Garcia-Castro\orcidlink{0000-0003-3379-4495}}
\affiliation{School of Physics, \href{https://ror.org/00xc1d948}{Universidad Industrial de Santander}, Carrera 27 Calle 9, Bucaramanga SAN-680002, Colombia}

\author{Oleg Rubel\orcidlink{0000-0001-5104-5602}}
\email{rubelo@mcmaster.ca}
\affiliation{Department of Materials Science and Engineering, \href{https://ror.org/02fa3aq29}{McMaster University}, 1280 Main Street West, Hamilton, Ontario L8S 4L8, Canada}

%Wuttig, Gonze, S Buegel, versstrate

\date{\today}% It is always \today, today,

\begin{abstract}
We performed high-throughput density functional theory calculations of optical matrix elements between band edges across a diverse set of non-magnetic two-dimensional monolayers with direct band gaps. Materials were ranked as potential optical emitters, leading to the identification of transition-metal nitrogen halides (ZrNCl, TiNBr, TiNCl) and bismuth chalcohalides (BiTeCl) with optical coupling comparable to or exceeding \ce{MoS2}. Despite strong in-plane dipole transitions, most two-dimensional materials underperform bulk semiconductors due to the absence of out-of-plane components. To elucidate interband transitions, we introduced the orbital overlap tensor and established a correlation between anomalous Born effective charges and optical coupling, linking charge redistribution to transition strength. We also identified chalcogen-mediated $d$-$d$ transition as a key mechanism enabling optical responses in transition-metal dichalcogenides. We derived an analytical radiative recombination model incorporating multi-valley effects and found that excitonic corrections are essential for accurate lifetime predictions. Some direct-gap materials exhibit dark excitons as their lowest-energy states, classifying them as quasi-direct band gap semiconductors, which is critical for tuning excitonic recombination dynamics.
\end{abstract}

%Use showkeys class 
\maketitle

%%%%%%%%%%%%%%%%%%%%%%%%%%%%%%%%%%%%%%%%%%%%%%%%%%%%%%%%%%%%%%%%%%%%%%%%%%%%%%%%%%%%%%%%%%%%%%%%%%%%%%%%%%%%%%%%%%%%%%%%%%%%%%%%%%%%%%%%%%%%%
\section{Introduction}\label{sec:Introduction}
%%%%%%%%%%%%%%%%%%%%%%%%%%%%%%%%%%%%%%%%%%%%%%%%%%%%%%%%%%%%%%%%%%%%%%%%%%%%%%%%%%%%%%%%%%%%%%%%%%%%%%%%%%%%%%%%%%%%%%%%%%%%%%%%%%%%%%%%%%%%%

The emergence of \acp{2DM} as physically realizable systems has resulted in one of the most active fields in modern materials research~\cite{Novoselov_RevModPhys_83_2011_https://doi.org/10.1103/revmodphys.83.837,Glavin_Adv.Mater._2020_https://doi.org/10.1002/adma.201904302,Butler_ACSNano_2013_7_https://doi.org/10.1021/nn400280c}. Atomically thin materials provide an ideal platform for investigating electronic properties under quantum confinement, particularly the enhancement of excitonic effects~\cite{Novoselov_RevModPhys_83_2011_https://doi.org/10.1103/revmodphys.83.837,Glavin_Adv.Mater._2020_https://doi.org/10.1002/adma.201904302,Butler_ACSNano_2013_7_https://doi.org/10.1021/nn400280c,Wang_RevModPhys.90.021001_10.1103/RevModPhys.90.021001}. Reducing layered materials to a single sheet opens up possibilities for modifying the optoelectronic properties of these materials, such as band gap engineering~\cite{Chaves_2DMatterApp_4_1_2020_https://doi.org/10.1038/s41699-020-00162-4}. Furthermore, the transition from a bulk material to a two-dimensional monolayer can lead to a crossover from an electronic indirect band gap to a direct band gap material, which is promising for light-emitting applications. Prominent examples of materials that exhibit this modification of band type include members of the \acp{TMDC} family~\cite{Ramasubramaniam_PhysRevBCondensMattPhys_86_2012_https://doi.org/10.1103/physrevb.86.115409,Gutierrez_NanoLett_8_2013_https://doi.org/10.1021/nl3026357}, tellurene~\cite{Sang_Nanomat_9_2019_https://doi.org/10.3390/nano9081075}, and phosphorene (black phosphorus)~\cite{Buscema_Nanolett_14_2014_https://doi.org/10.1021/nl5008085,Tran_PhysRevB_89_2014_https://doi.org/10.1103/physrevb.89.235319,Wang_NatNannotech_10_2015_https://doi.org/10.1038/nnano.2015.71}. Among the \ac{TMDC} family, the monolayer of molybdenum disulfide \ce{MoS2} has been extensively studied, and its electronic properties have been predicted through \textit{ab initio} simulations and confirmed by experiments~\cite{Ramasubramaniam_PhysRevBCondensMattPhys_86_2012_https://doi.org/10.1103/physrevb.86.115409,Mak_PhysRevLett_105_2010_https://doi.org/10.1103/physrevlett.105.136805,Mak_NatNanotechnol_7_8_2012_https://doi.org/10.1038/nnano.2012.96,Qiu_PhysRevLett.111.216805_10.1103/PhysRevLett.111.216805}. This 2D direct band gap semiconductor material exhibits strong light emission through luminescence~\cite{Sallen_PjysRevBCondensMatterPhys_86_2012_https://doi.org/10.1103/physrevb.86.081301,Splendian_NanoLett_10_2010_https://doi.org/10.1021/nl903868w}, high light absorption~\cite{Newaz_SolidStateCommun_155_2013_https://doi.org/10.1016/j.ssc.2012.11.010}, strongly bound exciton peaks~\cite{Ramasubramaniam_PhysRevBCondensMattPhys_86_2012_https://doi.org/10.1103/physrevb.86.115409}, and circular dichroism~\cite{Cao_NatCommun_3_2012_https://doi.org/10.1038/ncomms1882,Zeng_NatNanotechno_7_2012_https://doi.org/10.1038/nnano.2012.95}. As a result, there is a growing interest in studying the optical properties of 2D direct band gap materials~\cite{Wang_InfraredPhys_88_2018_https://doi.org/10.1016/j.infrared.2017.11.009,Wang_ChemSocRev_47_2018_https://doi.org/10.1039/c8cs00255j,Gao_J_Mater._9_4_2023_https://doi.org/10.1016/j.jmat.2023.02.005,Li_Proc_IEEEInst.Electr.Electron.Eng._2020_108_5_https://doi.org/10.1109/jproc.2019.2936424,Xie_AdvMater._33_22_2021_https://doi.org/10.1002/adma.201904306,Zhao_Nanophotonics_4_2015_https://doi.org/10.1515/nanoph-2014-0022,Bernardi_Nanophotonics_6_2_2017_10.1515/nanoph-2015-0030_2017}.

While the presence of a direct band gap is a necessary condition for efficient optical emitters, it is insufficient on its own. In solid-state systems, under the assumption of the electric dipole approximation, the inter-band optical matrix elements for unpolarized light are directly proportional to the linear momentum matrix elements, denoted as $\bm{p}_{vc,\bm{k}} = \langle \phi_{c,\bm{k}}| \hat{\bm{p}} |\phi_{v,\bm{k}}\rangle$. Here, $\hat{\bm{p}}$ represents the momentum operator, $\bm{k}$ is a wave vector within the \ac{BZ}, and $|\phi_{v,\bm{k}}\rangle$ and $|\phi_{c,\bm{k}}\rangle$ denote Bloch states in the occupied valence band and the unoccupied conduction band, respectively (see Refs.~\cite[chap.~3]{Fox_OxfortUniPress_2010} and~\cite[chap.~9]{Cohen_2019_Fundamentalsofcondensedmatterphysics}). The momentum matrix elements play a crucial role in studying the absorption and emission characteristics of direct band gap semiconductors~\cite{Masakatsu_JpnJApplPhys_35_2R_https://doi.org/10.1143/jjap.35.543,Shokhovets_AppPhysLett_86_16_2005_https://doi.org/10.1063/1.1906313,Rhim_PhysRevB_71_4_2005_https://doi.org/10.1103/physrevb.71.045202}, as they are essential for the computation of the optical absorption coefficient, radiative lifetime, and luminescence intensity~\cite[chap.~3 and 4]{Fox_OxfortUniPress_2010}. For bulk group-IV, III-V, and II-VI semiconductors, the momentum matrix elements are often expressed in energy units~\cite[p.~71]{Cardona_Springer_2016}
\begin{equation}\label{eq:E_P}
    E_P = \frac{2\langle p_{vc_1}^2\rangle_{\alpha}}{m_0},
\end{equation}
where $m_0$ is the rest mass of an electron and $\langle p_{vc_1}^2 \rangle_{\alpha}=(p_{x,vc_1}^2 + p_{y,vc_1}^2 + p_{z,vc_1}^2)/3$ is the average matrix element over Cartesian directions $\alpha=x,y,z$ to account for an arbitrary polarization of light. Here, $p_{vc_1}$ includes coupling between the heavy-hole, light-hole, split-off bands, and only \textit{one} of the double-degenerate \ac{CBE} states at $\Gamma$ point. The optical coupling between the band edges in these traditional optoelectronic materials typically falls in the range of $E_P = 14\!-\!31.4$~eV, exhibiting low sensitivity to chemical composition~\cite{Hermann_PhysRev_15_2_1977_https://doi.org/10.1103/physrevb.15.823,Pfeffer_PhysRevB_52_19_1996_https://doi.org/10.1103/physrevb.53.12813,Cardona_PhysRevB_38_3_1988_https://doi.org/10.1103/physrevb.38.1806,Vurgaftman_J.Appl.Phys_89_11_2001_https://doi.org/10.1063/1.1368156,Bernardi_Nanophotonics_6_2_2017_10.1515/nanoph-2015-0030_2017}. It is intriguing to compare this characteristic to the optical coupling in \acp{2DM}.

A substantial body of optical calculations related to \acp{2DM} has been conducted, focusing on properties such as absorption spectra and interband polarizability~\cite{Ketolainen_JCTC_16_2020_10.1021/acs.jctc.0c00387,Gupta_ACSNano_12_11_2018_https://doi.org/10.1021/acsnano.8b03754,Bernardi_NanoLett_13_8_2013_https://doi.org/10.1021/nl401544y,Ramasubramaniam_PhysRevBCondensMattPhys_86_2012_https://doi.org/10.1103/physrevb.86.115409,Qiu_PhysRevLett.111.216805_10.1103/PhysRevLett.111.216805,Wang_RevModPhys.90.021001_10.1103/RevModPhys.90.021001,Kumar_MaterChemPhys_135_2-3_2012_https://doi.org/10.1016/j.matchemphys.2012.05.055,Haastrup_2DMATER_5_2018_https://doi.org/10.1088/2053-1583/aacfc1, Gu2023_10.3390/nano13010196, Jameel2024_10.1007/s10904-023-02828-0}. However, the results of these calculations are not directly correlated with the efficiency of these materials as optical emitters, which requires knowledge of the optical coupling strength at the band edges. The momentum or velocity matrix elements at the band edge of monolayer \ce{MoS2}, as reported in Ref.~\citenum{Rubel_ComputationMPDI_110_2_2022_https://doi.org/10.3390/computation10020022} allow to evaluate the in-plane value of $E_P=4.1\!-\!5.8$~eV ($\alpha=x,y$) at the $K$-point of the hexagonal 2D \ac{BZ} (the uncertainty is due to the choice between theoretical approximations for the exchange-correlation functional). Thus, the optical coupling in one of the most studied \ac{2DM} falls short of traditional bulk semiconductors by a factor of approximately four. Nonetheless, there is still a lack of information about whether other \acp{2DM} have similar optical matrix elements, as these values have not been reported in the existing literature. 

The present study aims to fill this gap by investigating the momentum matrix elements of \acp{2DM} using \ac{DFT}~\cite{HohenbergKohn_Physrev_136_1964_10.1103/PhysRev.136.B864,KohnSham_Physrev_140_1965_10.1103/PhysRev.140.A1133} within a high-throughput framework. Structural data for the \acp{2DM} were sourced from the \ac{C2DB}~\cite{Haastrup_2DMATER_5_2018_https://doi.org/10.1088/2053-1583/aacfc1,Gjerding_2DMaterials_8_4_2021_https://dx.doi.org/10.1088/2053-1583/ac1059}. Our focus specifically encompassed thermodynamically and dynamically stable, non-magnetic, direct band gap semiconductors. Of the 15,733 materials in the database, we filtered the data according to the constraints explained in Sec.~\ref{sec:Computational and theoretical details}, ultimately calculating the momentum matrix elements between the \ac{VBE} and \ac{CBE} for 358 monolayers. We determined the orbital character of \ac{VBE} and \ac{CBE} states and categorized the materials based on the proportions of $s \to p$, $p \to d$, and forbidden ($\Delta \ell \neq \pm 1$) transitions. Counterintuitively, \acp{TMDC} predominantly exhibited forbidden (mainly $d \to d$) transitions, despite demonstrating significant optical activity. We attribute this observation to a high polarizability of transition metal non-bonding $d$-states. This phenomenon is observed in conjunction with a dynamic charge transfer and was first recognized by~\citet{Gonze_PhysRevB.95.201106_10.1103/PhysRevB.95.201106} in the context of anomalous \acp{BEC} in hexagonal \acp{TMDC}. We compared strength of the optical coupling at the band edges of the \acp{2DM} to those of conventional optoelectronic bulk semiconductors. To illustrate practical implications of momentum matrix elements at the band edges, we computed the radiative recombination coefficient $B_{\text{2D}}$ and the radiative lifetime, which is an experimentally accessible physical quantity. These parameters are critical for optoelectronic applications such as light-emitting diodes, lasers, and photosensors.

%%%%%%%%%%%%%%%%%%%%%%%%%%%%%%%%%%%%%%%%%%%%%%%%%%%%%%%%%%%%%%%%%%%%%%%%%%%%%%%%%%%%%%%%%%%%%%%%%%%%%%%%%%%%%%%%%%%%%%%%%%%%%%%%%%%%%%%%%%%%%
\section{Computational details}\label{sec:Computational and theoretical details}
%%%%%%%%%%%%%%%%%%%%%%%%%%%%%%%%%%%%%%%%%%%%%%%%%%%%%%%%%%%%%%%%%%%%%%%%%%%%%%%%%%%%%%%%%%%%%%%%%%%%%%%%%%%%%%%%%%%%%%%%%%%%%%%%%%%%%%%%%%%%%

The crystal structures utilized in this investigation were obtained from the \ac{C2DB} database~\cite{Haastrup_2DMATER_5_2018_https://doi.org/10.1088/2053-1583/aacfc1}. Two-dimensional semiconductors exhibiting direct electronic band gaps of up to 3.5~eV were selected based on their dynamic and thermodynamic stability. The selection criteria required non-imaginary frequencies in the phonon dispersion and energy values above the convex hull of less than 50~meV per atom. For the most promising materials (high $\langle p^{2}_{vc,\bm{k}_0} \rangle_{\alpha,T}$), discussed in this text, we conducted additional assessments of their viability by identifying a parent two-dimensional bulk structure that has been either experimentally reported or listed as stable in the Materials Project database~\cite{Jain_AM_1_2013, Ong_CM_20_2008}.

The \ac{DFT}~\cite{PhysRev.136.B864,PhysRev.140.A1133} based calculations were conducted using the \ac{VASP} package~\cite{Kresse_physrevb_47_1993_10.1103/PhysRevB.47.558,Kresse_compmatscience_6_1996_10.1016/0927-0256(96)00008-0} (version 6.4.0), which employs the projector augmented wave pseudopotential method for the basis set treatment~\cite{Bloch_physrevb_50_1994_10.1103/PhysRevB.50.17953}. The \ac{PBE} exchange-correlation functional was employed~\cite{Perdew_physrevlett_77_1996_10.1103/PhysRevLett.77.3865}. The cutoff energy for the plane wave expansion was set to the maximum \texttt{ENMAX} parameter (\texttt{POTCAR} file) from the atomic species in the compounds. Projector augmented-wave pseudopotentials~\cite{Kresse_PRB_59_1999} (version 5.4) were employed. The number of valence electrons and cutoff energy for each element can be found in Table~\ref{table:SI_Potcars_1} from the \ac{SM}. A $\Gamma$-centered unshifted $k$-mesh was utilized. We employed the fully automatic generation scheme with a $R_{k}=20$ parameter, which determines the number of subdivisions for every reciprocal lattice vector $\bm{b}_{i}$ as $N_{i} = \text{int}[\text{max}(1,R_{k}|\bm{b}_{i}|+0.5)]$. In cases where the $k$-point corresponding to the band edges was not included in the automatic mesh, it was explicitly added to the list with a weight of zero for the computation of the momentum matrix elements. The precision mode (\texttt{PREC} flag) was set to `accurate'. This resulted in a denser Fourier grid for charge densities and potentials. \Ac{SOC} was considered in all calculations~\cite{PhysRevB.93.224425}. The convergence criteria for the total energy in the electronic self-consistent field loop was set at $10^{-6}$~eV. For the orbital partial occupancies, we used a 0.01~eV Gaussian smearing width.

The momentum matrix elements were transformed from the dipole matrix elements $\bm{r}_{vc,\bm{k}}$, read from \ac{VASP} \texttt{WAVEDER} file, as
\begin{equation}\label{eq:r -> p}
    \bm{p}_{vc,\bm{k}} = 
    \frac{\mathbbm{i} m_0}{\hbar}(E_{c,\bm{k}} - E_{v,\bm{k}})  \bm{r}_{vc,\bm{k}},
\end{equation}
where $\mathbbm{i}$ is the imaginary unit, $\hbar$ is the reduced Planck constant, and $E_{v/c,\bm{k}}$ are \ac{DFT} energy eigenvalues of states in the valence/conduction band. Convergence tests for the absolute square momentum matrix elements were performed on three representative materials with respect to $R_{k}$, cutoff energy, and precision mode computational parameters. Results of the tests are presented in Fig.~\ref{sup:fig-convergence} of the \ac{SM}. It is noteworthy that both the oscillator strength and the dipole matrix elements are influenced by the band gap value via the term $E_{c,\bm{k}} - E_{v,\bm{k}}$ in Eq.~\eqref{eq:r -> p}. However, accurately determining the band gap in standard \ac{DFT} calculations can be problematic, making it challenging to compare values of the oscillator strength and of the dipole matrix elements at the band edge between different materials. On the other hand, $\bm{p}_{vc,\bm{k}}$ is not affected by the band gap error, providing a more suitable metric for comparing materials~\cite{Laurien_PhysRevB_106_4_2022_https://doi.org/10.1103/physrevb.106.045204},~\cite[app.~A]{Blood2015_Oxforf_2015}.

Effective charges (Bader and Born) were sourced from the \ac{C2DB} database~\cite{Haastrup_2DMATER_5_2018_https://doi.org/10.1088/2053-1583/aacfc1,Gjerding_2DMaterials_8_4_2021_https://dx.doi.org/10.1088/2053-1583/ac1059}. For instances without reported Bader charabsolute squareges, we calculated them and verified their convergence (Tables~\ref{table:SI:Bader} and \ref{table:SI:Bader_conver} from the \ac{SM}). The cell-periodic part of pseudo-wavefunctions $\Tilde{u}_{\bm{k}}(\bm{r})$ was extracted using VASPKIT~\cite{Wang_2021_https://doi.org/10.1016/j.cpc.2021.108033} to visualize the overlap $ \Tilde{u}_{\bm{k}}^{*}(\bm{r}) \Tilde{u}_{\bm{k}+\bm{q}_{\alpha}}(\bm{r})$ in real space with a small shift $q$ of the magnitude approximately 0.004~{\AA}$^{-1}$. The bimolecular radiative recombination coefficients were computed at the \ac{DFT} level by adapting the methodology outlined by~\citet{Xu_CPC_305_2024_10.1016/j.cpc.2024.109352} to \acp{2DM}. The $k$-meshes employed for the calculation included at least a hundred times more $k$-points for the irreducible \ac{BZ} sampling than those employed for the momentum matrix elements calculation.

Excitonic properties of selected \acp{2DM} were calculated using \ac{VASP}'s implementation of the \ac{BSE}~\cite{Sander_PRB_92_2015} with the Tamm-Dancoff approximation. \Ac{SOC} was included for all materials. Preliminary ground state calculations, virtual orbital calculations, and derivatives of the orbitals with respect to the Bloch vectors were performed at the \ac{PBE} level. $\Gamma$-centered k-mesh with 32 divisors per {\AA}$^{-1}$ length of reciprocal lattice vectors is used to sample the \ac{BZ}. Quasiparticle energies and the screened Coulomb kernel were obtained from a subsequent $G_0 W_0$ calculation. At least 16 occupied and 16 unoccupied orbitals near the band edges were included in the \ac{BSE} calculation. Sample input files are available from a Zenodo repository~\cite{Zenodo_10.5281/zenodo.13773214}.
%%%%%%%%%%%%%%%%%%%%%%%%%%%%%%%%%%%%%%%%%%%%%%%%%%%%%%%%%%%%%%%%%%%%%%%%%%%%%%%%%%%%%%%%%%%%%%%%%%%%%%%%%%%%%%%%%%%%%%%%%%%%%%%%%%%%%%%%%%%%%
\section{Results and discussion}\label{sec:Results and discussion}
%%%%%%%%%%%%%%%%%%%%%%%%%%%%%%%%%%%%%%%%%%%%%%%%%%%%%%%%%%%%%%%%%%%%%%%%%%%%%%%%%%%%%%%%%%%%%%%%%%%%%%%%%%%%%%%%%%%%%%%%%%%%%%%%%%%%%%%%%%%%%

\subsection{Momentum matrix elements}

\begin{figure}
    \centering
    \includegraphics[width=6cm,keepaspectratio=true]{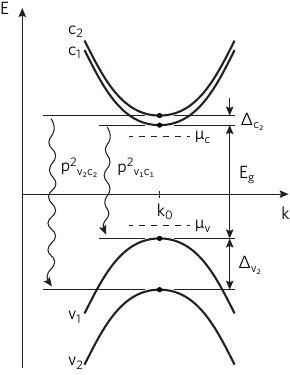}
    \caption{Schematic band structure with non-degenerate bands showing two optically active transitions from a total of four possible ones at the $\bm{k}_0$ valley. Band splittings $\Delta_{c_i}$ and $\Delta_{v_j}$ are positively defined and are measured relative to the band edges.}
    \label{fig-band-schematic}
\end{figure}

As an initial step, we have computed the momentum matrix elements $\bm{p}_{vc,\bm{k}}$ between states at the band edges based on the electronic structure of \acp{2DM}. These matrix elements are essential for understanding optical emission resulting from the recombination of charge carriers between \ac{CBE} and \ac{VBE}. Our study specifically focuses on direct band gap \acp{2DM}.  Generally, there may exist a group of bands $v_j \in \{v_1,v_2,\ldots\}$ and $c_i\in \{c_1,c_2,\ldots\}$ close to the band edge at $\bm{k}_0$ as illustrated in Fig.~\ref{fig-band-schematic}. These bands may either be degenerate, as in the case of monolayer \ce{ReS2}, or the degeneracies may be lifted, for example, due to \ac{SOC} as seen in monolayer \ce{MoS2}. The effective optical matrix element for spontaneous emission is obtained by summing over all possible pairs $v_j c_i$ as follows:
\begin{equation}\label{eq:p_vc average}
        \langle p^{2}_{vc,\bm{k}_0} \rangle_{\alpha,T} = \sum_{i,j} \langle p^2_{v_{j}c_{i},\bm{k}_0} \rangle_{\alpha} \exp[-(\Delta_{c_i} + \Delta_{v_j})/k_{\text{B}}T] ,
\end{equation}
where $\Delta_{c_i}$ and $\Delta_{v_j}$ are positively defined band splittings measured relative to the band edges (Fig.~\ref{fig-band-schematic}). The exponential term represents the probability of occupancy of energy levels above the band gap, under the assumption of a dilute limit for the charge carrier density. For instance, the lowest energy transition `A' dominates the photoluminescence spectrum of monolayer \ce{MoS2} at room temperature~\cite{Kaplan_2DM_3_2016} in contrast to the second lowest energy transition `B' with an excess energy of $\Delta_{c_2} + \Delta_{v_2} \approx0.15$~eV, which exhibits a considerably lower spectral intensity. Although the matrix elements $p^{2}_{v_1 c_1,K}$ and $p^{2}_{v_2 c_2,K}$ possess identical magnitudes, the contribution of the second transition to radiative recombination at room temperature is significantly  reduced owing to $\Delta_{c_2} + \Delta_{v_2} \gg k_{\text{B}}T$.

\begin{figure}
    \centering
    \includegraphics[width=9cm,keepaspectratio=true]{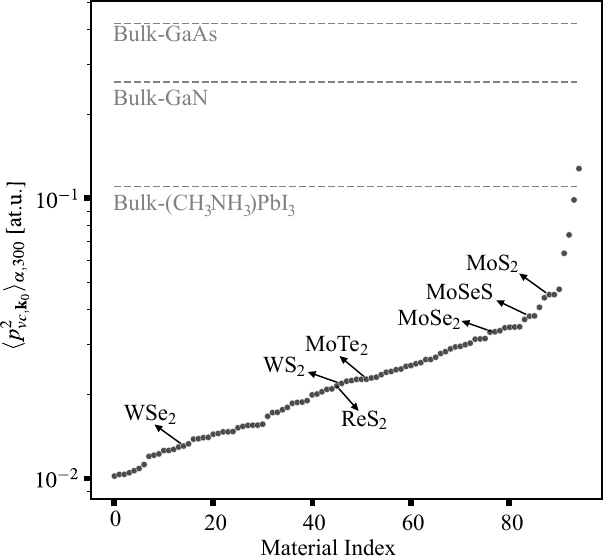}
    \caption{Ranking of direct band gap \acp{2DM} according to their optical activity captured by the polarization-average momentum matrix element at the band edges $\langle p_{vc}^2 \rangle_{\alpha,300~\text{K}}$  (in atomic units) defined by Eq.~\eqref{eq:p_vc average}. Only materials with the highest optical activity are presented and arranged in ascending order. The dashed lines mark values corresponding to the well-known optically-active bulk semiconductors.} 
    \label{fig:variability}
\end{figure}

Figure~\ref{fig:variability} presents ranking of \acp{2DM} according to the magnitude of momentum matrix elements computed using Eq.~\eqref{eq:p_vc average} and truncated at $\langle p^{2}_{vc,\bm{k}_0} \rangle_{\alpha,300~\text{K}}=0.01$~at.u. The optical activity of most \acp{TMDC} is ranked as above average. The following prominent \acp{2DM}, that have been previously synthesized, are labeled in Fig.~\ref{fig:variability}: \ce{MoS2}~\cite{Mak_PhysRevLett_105_2010_https://doi.org/10.1103/physrevlett.105.136805,Splendian_NanoLett_10_2010_https://doi.org/10.1021/nl903868w,Kappera_Nat.matter_13_12_https://doi.org/10.1038/nmat4080_2014,Lopez-Sanchez_Nat_Nanotechnol_8_2013_https://doi.org/10.1038/nnano.2013.100}, \ce{MoSeS}~\cite{Lu_Nat.Nanotechnol_12_8_https://doi.org/10.1038/nnano.2017.100_2017}, \ce{MoSe2} and \ce{WSe2}~\cite{Tonndorf_OptExpress_4_21_10.1364/OE.21.004908_2013},  \ce{MoTe2}~\cite{Wang_Nature_550_7677_https://doi.org/10.1038/nature24043_2017}, \ce{ReS2}~\cite{Tongay_Nat.Commun_5_1_https://doi.org/10.1038/ncomms4252_2014} and \ce{WS2}~\cite{Okada_ACS.Nano_8_8_https://doi.org/10.1021/nn503093k_2014}. The top ranked monolayer materials in Fig.~\ref{fig:variability} (not explicitly labeled) are derived from naturally occurring bulk layered phases, suggesting that exfoliation is plausible. Materials with the highest optical coupling between band edges are listed in Table~\ref{table:best_materials} along with well-established \acp{TMDC}. The optical coupling strength of the non-\acp{TMDC} materials is comparable and even superior to that of \acp{TMDC}, indicating potential for optoelectronic applications. Some of those materials have the band gap at the \ac{HSE06} level, including \ac{SOC}, within the visible range of the electromagnetic spectrum (1.8$-$3.1~eV). The direct transition occurs at $\Gamma$ point in the \ac{BZ} for all the non-\acp{TMDC} and at $K=(1/3, 1/3, 0)$ for the \acp{TMDC}.

\begin{table}[]
    \singlespacing
    \caption{Characteristics of direct band gap \acp{2DM} having strongest optical coupling at the band edges. The band gap $E_\text{g}$ and \ac{BSE} exciton energies $E_\text{exc}$ include \ac{SOC}. The superscript $(^{\times n})$ reflects the degeneracy of exciton states.}\label{table:best_materials}
    \begin{ruledtabular}
    \begin{tabular}{lccccccc}
        Material & $E_\text{g}$ & $\bm{k}_0$ & $\langle p_{vc,\bm{k}_0}^2 \rangle_{\alpha,300~\text{K}}$ & $E_\text{exc}^{(\lambda)}$ & $\langle p_{\text{exc}}^2(\lambda) \rangle_{\alpha}$ & \ac{C2DB} id & Parent bulk   \\
        & (eV)\footnotemark[1] & & (at. u.) & (eV) & (at. u.) & & structure\footnotemark[2] \\
        \hline
        GaSeCl & 3.6 & $\Gamma$ & 0.13 & \begin{tabular}[t]{@{}c@{}}3.60\\3.61\\3.61\\3.66\end{tabular} & \begin{tabular}[t]{@{}c@{}}0\\0.01\\0\\0.59\end{tabular} & 2ClGaSe-1 &~\cite{Kniep_MRB_18_1983}, mp-1120728 \\
        BiTeCl & 1.2 & $\Gamma$ & 0.10 & \begin{tabular}[t]{@{}c@{}}$1.33^{\times 2}$\\$1.34^{\times 2}$\end{tabular} & \begin{tabular}[t]{@{}c@{}}$0.049^{\times 2}$\\$0.016^{\times 2}$\end{tabular} & 1BiClTe-1 &  mp-28944  \\
        \ce{Bi2Se2Te} & 0.76 & $\Gamma$ & 0.074 & $0.70^{\times 4}$ & \begin{tabular}[t]{@{}c@{}}$0^{\times 2}$\\$0.039^{\times 2}$\end{tabular} & 1TeBi2Se2-1 & \begin{tabular}[t]{@{}c@{}}sd\_1727317\\ cod-9004849\end{tabular} \\
        ZrNCl & 3.1 & $\Gamma$ & 0.064 & \begin{tabular}[t]{@{}c@{}}$2.50^{\times 3}$\\$2.51$\end{tabular} & \begin{tabular}[t]{@{}c@{}}$0^{\times 3}$\\$0.13$\end{tabular} & 2ClNZr-2 & sd\_1704456 \\
        TiNBr & 2.1 & $\Gamma$ & 0.047 & $1.34^{\times 4}$ & \begin{tabular}[t]{@{}c@{}}$0^{\times 3}$\\$0.053$\end{tabular} & 2BrNTi-1 & sd\_1704458 \\
        \ce{MoS2} & 2.1 & $K$ & 0.045 & $1.96^{\times 4}$ & \begin{tabular}[t]{@{}c@{}}$0^{\times 2}$\\$0.073^{\times 2}$\end{tabular} & 1MoS2-1 & sd\_0309036 \\
        TiNCl & 2.1 & $\Gamma$ & 0.044 & $1.37^{\times 4}$ & \begin{tabular}[t]{@{}c@{}}$0^{\times 3}$\\0.052\end{tabular} & 2ClNTi-1 & sd\_1704459 \\
        \ce{HgI2} & 2.5 & $\Gamma$ & 0.041 & \begin{tabular}[t]{@{}c@{}}2.51\\2.52\\$2.57^{\times 2}$\\2.62\\$2.67^{\times 2}$\end{tabular} & \begin{tabular}[t]{@{}c@{}}0\\0\\$0.024^{\times 2}$\\0\\$0.13^{\times 2}$\end{tabular}& 4HgI2-1 & cod-9008155 \\
        MoSeS & 1.9 & $K$ & 0.038 & $1.82^{\times 4}$ & \begin{tabular}[t]{@{}c@{}}$0.067^{\times 2}$\\$0^{\times 2}$\end{tabular} & 1MoSSe-1 &~\cite{Lu_Nat.Nanotechnol_12_8_https://doi.org/10.1038/nnano.2017.100_2017}, mp-1221404 \\
        \ce{MoSe2} & 1.8 & $K$ & 0.033 & \begin{tabular}[t]{@{}c@{}}$1.67^{\times 2}$\\$1.68^{\times 2}$\end{tabular} & \begin{tabular}[t]{@{}c@{}}$0.060^{\times 2}$\\$0^{\times 2}$\end{tabular} & 1MoSe2-1 & sd\_0309034\\
        \hline
        \ce{Mg2Al2Se5} & 2.0 & $\Gamma$ & 0.013 & \begin{tabular}[t]{@{}c@{}}$2.13^{\times 2}$\\$2.14^{\times 2}$\end{tabular} & \begin{tabular}[t]{@{}c@{}}$0^{\times 2}$\\$0.011^{\times 2}$\end{tabular} & 1Al2Mg2Se5-1 & mp-29624
    \end{tabular}
    \end{ruledtabular}
    \footnotetext[1]{\noindent Data from \ac{C2DB} calculated at the Heyd-Scuseria-Ernzerhof'06 screened hybrid functional~\cite{Krukau_2006_JChemPhys125_22_https://doi.org/10.1063/1.2404663} level including \ac{SOC}.\hfill~}
    \footnotetext[2]{The prefix in structures id's corresponds to the following databases: `mp' Materials Project~\cite{Jain_AM_1_2013, Ong_CM_20_2008}, `cod' Crystallography Open Database~\cite{Vaitkus_JC_15_2023}, `sd' Springer Materials.\hfill~}
\end{table}

The dashed lines in Fig.~\ref{fig:variability} represent the optical matrix elements for selected bulk semiconductors. The momentum matrix elements $\langle p^2_{v_{j}c_{i},\bm{k}_0} \rangle_{\alpha}$ for individual transitions were extracted from Ref.~\citenum{Rubel_ComputationMPDI_110_2_2022_https://doi.org/10.3390/computation10020022} at the \ac{DFT}-\ac{PBE} level and incorporated into Eq.~\eqref{eq:p_vc average} to compute the effective optical matrix elements between band edges. Comparing optical coupling strengths reveals that the best-performing \acp{2DM} fall short of established bulk optoelectronic materials. The primary reason for the reduced optical performance of \acp{2DM} is the lack of optical coupling with out-of-plane light polarization in most monolayer materials. Particularly, the absence of out-of-plane mirror symmetry is a necessary but not sufficient condition for the existence of $p_{z,v_j c_i}^2$ components. Additional factors limiting radiative recombination in \acp{TMDC} include spin-forbidden $K_{v_1}\!-\!K_{c_2}$ transitions and the underpopulation of the $v_2$ band in the $K$ valley due to the large band offset $\Delta_{\text{so}}$~\cite{Zhu_10.1103/PhysRevB.84.153402,Cheiwchanchamnangij_10.1103/PhysRevB.85.205302,Qiu_PhysRevLett.111.216805_10.1103/PhysRevLett.111.216805}.

\subsection{Orbital character of band edges and \acp{TMDC} anomaly}

Analysis of the orbital composition of states involved in radiative transitions will provide insight into the factors influencing the strength variability of optical transitions.  Vectors of orbital characters associated with states in the vicinity of \ac{VBE} and \ac{CBE} are expressed as 
\begin{subequations}\label{eq:vectors}
    \begin{align}
        \bm{v} & = (s_{v}, p_{v}, d_{v}) =
        \left(
            \sum_j s_{v_j} \exp[-\Delta_{v_j}/k_{\text{B}}T],
            \sum_j p_{v_j} \exp[-\Delta_{v_j}/k_{\text{B}}T],
            \sum_j d_{v_j} \exp[-\Delta_{v_j}/k_{\text{B}}T]
        \right) \label{eq:vectors v}\\
        \bm{c} & = (s_{c}, p_{c}, d_{c}) =
        \left(
            \sum_i s_{c_i} \exp[-\Delta_{c_i}/k_{\text{B}}T],
            \sum_i p_{c_i} \exp[-\Delta_{c_i}/k_{\text{B}}T],
            \sum_i d_{c_i} \exp[-\Delta_{c_i}/k_{\text{B}}T]
        \right)\label{eq:vectors c}
    \end{align}
\end{subequations}
with the summation indices $i$ and $j$ running over a set of conduction and valence bands, respectively. Symbols $s$, $p$, and $d$ represent the projected wavefunction character of each orbital, summed over all atomic sites (not to be confused with the notation for the momentum matrix elements, $p_{vc}$). The orbital characters of individual bands are weighted by the exponential factor (see schematic Fig.~\ref{fig-band-schematic}) to account for their occupancy at the finite temperature. Building upon the general idea of~\citet{Woods_Matter_6_9_https://doi.org/10.1016/j.matt.2023.06.043_2023}, we define a new metric (tensor) to characterize the orbital overlap between \ac{VBE} and \ac{CBE} states
\begin{equation}\label{eq:tensor product}
    \bm{P}'
    = \bm{v}\otimes\bm{c}
    =
    \begin{pmatrix}
        s_{v}s_{c} & s_{v}p_{c} & s_{v}d_{c}\\
        p_{v}s_{c} & p_{v}p_{c} & p_{v}d_{c}\\
        d_{v}s_{c} & d_{v}p_{c} & d_{v}d_{c}
    \end{pmatrix} .
\end{equation}
The matrix is normalized as
\begin{equation}\label{eq:tensor product norm}
    \bm{P} = \bm{P}'
    \left(
        \sum_{n,m=1}^{3} P'_{nm}
    \right)^{-1} ,
\end{equation}
resulting in $\sum_{n,m=1}^{3} P_{nm}=1$. Elements of the tensor can be assigned to transitions between orbitals as
\begin{subequations}\label{eq:transitions P}
    \begin{align}
        & P_{s \to p} = P_{12}+P_{21} , & \text{(allowed)}\\
        & P_{p \to d} = P_{23}+P_{32} , & \text{(allowed)}\\
        & P_{\times} = P_{11}+P_{22}+P_{33}+P_{13}+P_{31} . & \text{(forbidden)}
    \end{align}
\end{subequations}
The interpretation of transitions is based on (partial) orbital selection rules in atomic dipolar transitions \cite[chap.~4.3]{Fox2006_book_Quantumoptics}, which is a distant analogy to solids with dispersive bands.

\begin{figure}
    \centering
\includegraphics[width=13cm,keepaspectratio=true]{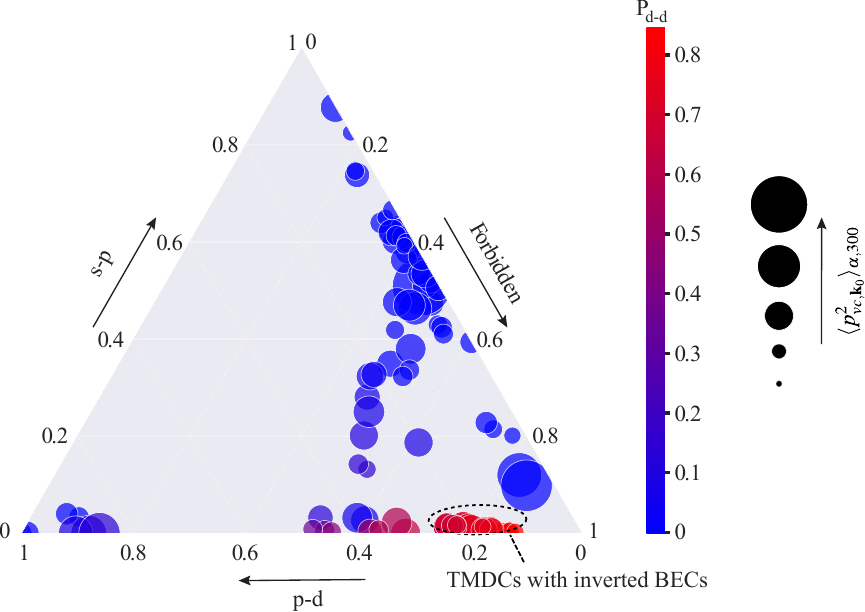}
    \caption{Orbital overlap ternary diagram where each \ac{2DM} from Fig.~\ref{fig:variability} is presented according to the fraction of allowed ($s \to p$, $p \to d$) and forbidden overlaps at the band edges, as defined by Eq.~\eqref{eq:transitions P} at room temperature. The size of the markers is proportional to the effective momentum matrix element at the band edges $\langle p_{vc}^2 \rangle_{\alpha,300~\text{K}}$ defined by Eq.~\eqref{eq:p_vc average}. The color refers to the fraction of $d \to d$ orbital overlap $P_{33}$.}
    \label{fig-thernary-d-d}
\end{figure}

Figure~\ref{fig-thernary-d-d} presents a ternary diagram in which all \acp{2DM} are positioned according to the fraction of allowed ($s \to p$, $p \to d$) and forbidden overlaps. Notably, there exists a considerable number of materials that exhibit relatively high optical coupling at the band edges alongside a significant weight of forbidden transitions. \acp{TMDC} are among those materials that demonstrate a high fraction of $d \to d$ transitions. In \acp{TMDC}, such as \ce{MoS2}, the electronic band edges are predominantly composed of $d$-orbitals of the Mo atom, with a small contribution from the $p$-orbitals of sulfur. Specifically, in the valence band at the $K = (1/3, 1/3, 0)$ point, Mo: $d_{x^2-y^2}$ and $d_{xy}$ orbitals are present, along with a small contribution from the sulfur $p_{x}$ and $p_{y}$ orbitals. In the conduction band at the $K$ point, the dominant contribution comes from the Mo $d_{z^2}$ orbital, which is consistent with prior findings~\cite{Xiao_PhysRevLett_108_19_10.1103/PhysRevLett.108.196802,Kormanyos2015_2dmatter_2_2_https://doi.org/10.1088/2053-1583/2/2/022001_2015}. According to atomic selection rules, this orbital composition is not expected to result in significant optical coupling. In contrast, hexagonal \acp{TMDC} exhibit some of the strongest matrix elements among \acp{2DM} (Fig.~\ref{fig:variability}).

Two explanations for the observed interband optical coupling in \acp{TMDC} have been proposed in the literature. The first proposed explanation pertains to \ac{SOC}, which is believed to facilitate the dipole transitions~\cite{Bernardi_NanoLett_13_8_2013_https://doi.org/10.1021/nl401544y,Bernardi_Nanophotonics_6_2_2017_10.1515/nanoph-2015-0030_2017}. However, our calculations of \ce{MoS2}, conducted without considering \ac{SOC}, indicated its effect on the optical matrix elements to be negligible. The second proposed explanation is the trigonal-prismatic crystal field splitting of Mo $d$-states ~\cite{Huisman_JournalSolidStateChem_3_1_1971_https://doi.org/10.1016/0022-4596(71)90007-7}. To test this idea, we performed a \ac{DFT} calculation of an isolated Mo atom in the trigonal-prismatic coordination of six S atoms in vacuum. The results confirmed the crystal field splitting for the $d$-electronic levels of Mo, with a lower energy level $A'_1$ composed of $d_{z^{2}}$ followed by two degenerate energy levels $E'$ composed of $d_{x^{2}-y^{2}}$ and $d_{xy}$, and a twofold degenerate higher energy levels $E''$ composed of $d_{xz}$ and $d_{yz}$. However, no optical coupling between $E'$ and $A'_1$ states was observed contrary to the claims in Ref.~\citenum{Huisman_JournalSolidStateChem_3_1_1971_https://doi.org/10.1016/0022-4596(71)90007-7}.

\begin{figure}%[H]
    \centering
\includegraphics[width=14.5cm,keepaspectratio=true]{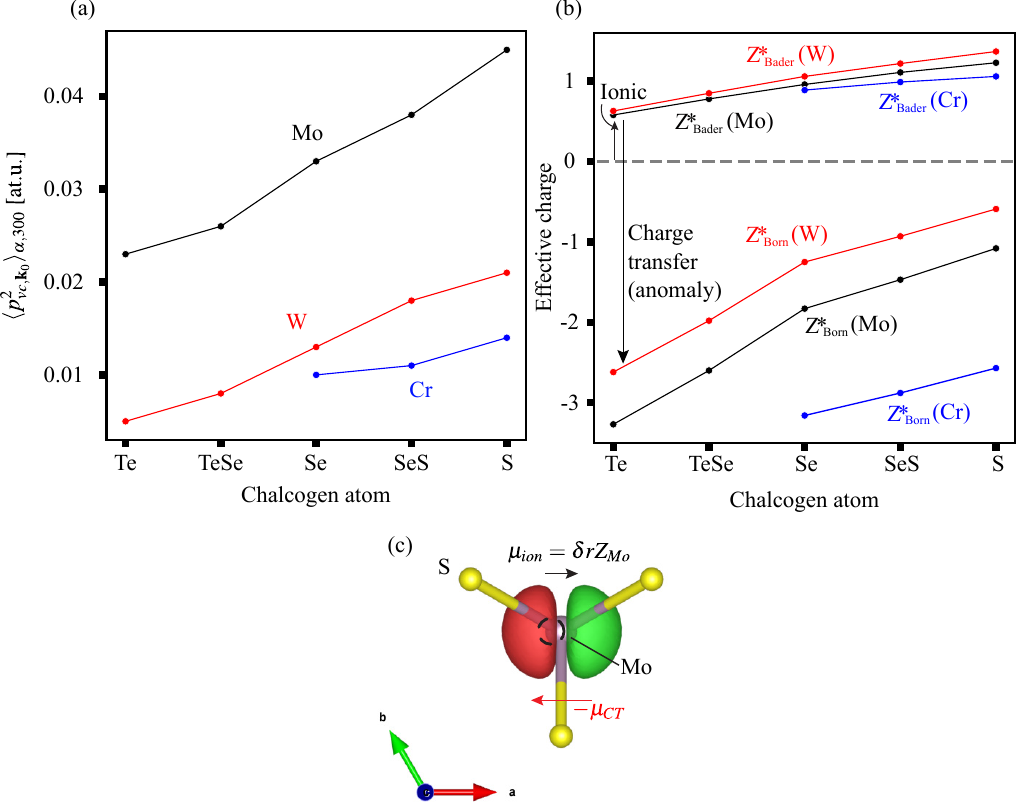}
    \caption{(a) Momentum matrix elements $\langle p_{vc}^2 \rangle_{\alpha,300~\text{K}}$  and (b) effective charges of the \ac{TM} atom plotted as a function of the chalcogen element in group-VI hexagonal \acp{TMDC}. The dynamical charge information is represented by the average of the $Z^{A}_{xx}$ and $Z^{A}_{yy}$ in-plane components of the Born effective charge tensor. The lines serve as visual aids. (c) Competing dipole moments contributing to the dynamic charge of Mo (see text for details). The local charge transfer (red for charge augmentation and green for charge depletion) gives rise to anomalous \acp{BEC}.}
    \label{fig:tmdcs}
\end{figure}

To elucidate the efficiency of optical coupling at the band edge, we analyzed chemical trends within the family of group-VI \acp{TMDC}, specifically \ce{$MX$2} (space group no.~187) and \ce{$MZX$} (space group no.~156), where $M$ represents Mo, W, or Cr, and $X$ and $Z$ denote S, Se, or Te. The optical coupling values exhibit an increase in materials with more electronegative chalcogen atoms, as illustrated in Fig.~\ref{fig:tmdcs}(a). To further investigate this chemical trend, we examined the Bader charges, which estimate charge transfer and reveal electron density redistribution due to bonding. Consequently, they are often used to infer the oxidation states of elements in solids. The values for the Bader charges depicted in Fig.~\ref{fig:tmdcs}(b) indicate that electrons transfer from the \ac{TM} to the chalcogen atom, suggesting an ionic character of bonding. This observed trend deviates from the conventional understanding in the literature, where enhanced overlap between band edge states (orbital hybridization), rather than electron transfer, is considered the driving mechanism for increased optical coupling~\cite{Kato_PhysRevMat_4_2020_https://doi.org/10.1103/physrevmaterials.4.035402, Maier_AdvMater._32_49_2020_https://doi.org/10.1002/adma.202005533, Wuttig_AdvFuncMat_32_2022_https://doi.org/10.1002/adfm.202110166, Schon_SciAdv_8_47_2022_https://doi.org/10.1126/sciadv.ade0828, Welnic_PhysRevLett_98_23_2007_https://doi.org/10.1103/physrevlett.98.236403}. 

We analyze the electronic charge redistribution in response to atomic displacements, which manifests as induced macroscopic polarization and is captured by the \acp{BEC}~\cite{Gonze_PhysRevB.55.10355}, to further clarify the counterintuitive correlation between bonding mechanisms and optical coupling. The \acp{BEC} characterize the dynamical behavior of charge and take into account contribution from both ionic charge and electron density redistribution in response to the perturbation~\cite{Gonze_PhysRevB.55.10355}. They have been extensively employed in the study of ferroelectric materials~\cite{Ghosez_PhysRevB.58.6224}. It has been established that the presence of \acs{BEC} magnitudes that significantly deviate from nominal ionic charges is attributed to dynamical contributions, which arise from two charge redistribution mechanisms. The first mechanism involves the delocalized transfer of electrons or interatomic hybridization, while the second mechanism pertains to on-site changes in hybridization or local displacements of the electronic cloud relative to the atom~\cite{Ghosez_PhysRevB.58.6224}.
 
In the case of the examined \acp{TMDC} monolayers, the Bader static and Born dynamic charges exhibit opposite sign (Fig.~\ref{fig:tmdcs}b). The entire set of monolayers with inverted charges are circled in Fig.~\ref{fig-thernary-d-d}. This phenomenon was explained by~\citet{Gonze_PhysRevB.95.201106_10.1103/PhysRevB.95.201106}, who found that an applied external electric field induces electron density changes predominantly localized around Mo atoms, exhibiting a hybridized $d$-orbital configuration. In contrast, the alteration in electron density surrounding S atoms is negligible. These non-bonding states, located at the top of the valence band, are sensitive to displacements of the Mo atoms, leading to local changes in polarization near the Mo sites (Fig.~\ref{fig:tmdcs}c). The dipole $\mu_{CT}$ resulting from the charge transfer near Mo atom counteracts the ionic component $\delta r Z_{\text{Mo}}$. This phenomenon governs the sign inversion of \acp{BEC} in hexagonal \acp{TMDC}~\cite{Gonze_PhysRevB.95.201106_10.1103/PhysRevB.95.201106}. This behavior establishes a heuristic link between anomalous \acp{BEC} and interband optical transitions. Consequently, the high optical coupling in \acp{TMDC} can be linked to the high polarizability of \ac{TM} $d$-states particularly at the band edge. The chalcogen $p$-orbitals mediate interaction between \ac{TM} $d$-orbitals without directly contributing to the band edge states.

\begin{figure}
    \centering
    \includegraphics{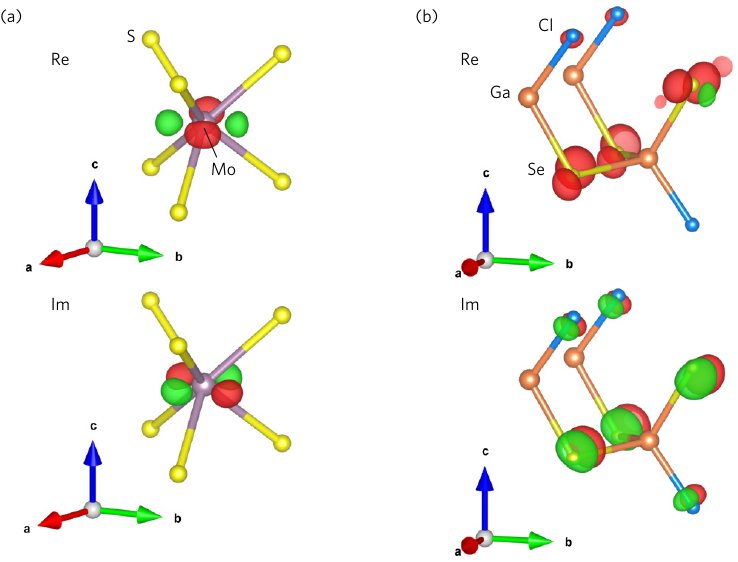}
    \caption{Real-space overlap $\Tilde{u}_{v,\bm{k}}^*(\bm{r}) \, \Tilde{u}_{c,\bm{k}+\bm{q}_{\alpha}}(\bm{r})$, representing the interband optical matrix element induced by an electromagnetic field polarized along Cartesian direction $\alpha$. (a) \ce{MoS2} at the $K$ point, direction $y$. (b) GaSeCl at the $\Gamma$ point, direction $x$. Isosurfaces show the real and imaginary parts of the overlap, with red and green indicating positive and negative values, respectively. The phase is chosen such that $\Im[\langle \Tilde{u}_{v,\bm{k}} | \Tilde{u}_{c,\bm{k}+\bm{q}_{\alpha}}\rangle]=0$.}
    \label{fig:overlap-real-space}
\end{figure}

To demonstrate that the optical activity in \acp{TMDC} is localized around the \ac{TM} sites for transitions between band edges, we use the correspondence between velocity and momentum matrix elements, $\bm{p}=m_0\bm{v}$. This relation is valid for local potentials~\cite{Starace_PRA_3_1971}, such as \ac{PBE} in \ac{DFT}. The velocity matrix element can be expressed in terms of an overlap matrix element between two states with a small finite displacement $q$ in reciprocal space~\cite{Rohlfing_PRB_62_2000, Rhim_PRB_71_2005, Rubel_ComputationMPDI_110_2_2022_https://doi.org/10.3390/computation10020022}
\begin{equation}
        v_{\alpha,vc}(\bm{k}) 
    = \lim_{q \to 0} \, (\hbar q)^{-1}
    \langle
        u_{c,\bm{k}+\bm{q}_{\alpha}} | u_{v,\bm{k}}
    \rangle
    \left[
        E_c(\bm{k}+\bm{q}_{\alpha}) - E_v(\bm{k})
    \right],
\end{equation}
where $u_{v/c,\bm{k}}(\bm{r})$ represents the cell-periodic part of the Bloch function for a state in the valence/conduction band.  For visualization, we employ the cell-periodic part of the Bloch pseudo-wavefunction $\Tilde{u}_{v/c,\bm{k}}(\bm{r})$ within the projector augmented wave formalism to evaluate the overlap $\Tilde{u}_{c,\bm{k}+\bm{q}_{\alpha}}^*(\bm{r})  \,\Tilde{u}_{v,\bm{k}}(\bm{r})$, serving as a measure of optical activity in real space. Figure~\ref{fig:overlap-real-space}a illustrates that the optical activity between band edges in monolayer \ce{MoS2} is indeed localized near the Mo sites within the Mo plane, with negligible involvement of S atoms. This behavior is strikingly different from materials like GaSeCl (layered group no.~32), where the high optical matrix element arises from the contributions of both Se and Cl atoms (Fig.~\ref{fig:overlap-real-space}b).

\subsection{Radiative recombination coefficient $B$ of selected compounds}

Next we will illustrate how optical matrix elements propagate into tangible optical characteristics of \acp{2DM}. The dynamics of radiative recombination of charge carriers within the bulk of an intrinsic semiconductor $(n = p)$ is governed by the equation
\begin{equation}
    \left(
        \frac{dn}{dt}
    \right)_{\text{rad}} = -B n^2
\end{equation}
with the bimolecular radiative recombination coefficient expressed as~\cite{Azarhoosh_AM_4_2016_10.1063/1.4955028, Xu_CPC_305_2024_10.1016/j.cpc.2024.109352}
\begin{equation}\label{eq:B(A)}
    B = (2\pi)^{-d} n^{-2} \sum_{i,j} \int_{\text{BZ}} f_{c_i,\bm{k}}(1-f_{v_j,\bm{k}})A_{v_j c_i,\bm{k}} ~d\bm{k},
\end{equation}
where $f$ is the Fermi-Dirac occupation probability, $A_{v_j c_i,\bm{k}}$ is the transition rate between states, and $d$ is a dimensionality of the system ($d=3$ in bulk or $d=2$ for \acp{2DM}). The integration can also be done over the irreducible \ac{BZ} by taking into account multiplicity of $k$ points. The transition rate between two states is given by the Einstein coefficient~\cite[p.~347]{Fox_OxfortUniPress_2010}
\begin{equation}\label{eq:A}
    A_{v_j c_i,\bm{k}} =
    \frac{
        n_{\text{r}} q_{\text{e}}^2
    }{
        \pi \epsilon_{0} \hbar^{2} m_{0}^{2} c^{3}
    } (E_{c_i,\bm{k}}-E_{v_j,\bm{k}}) \langle p^2_{v_j c_i,\bm{k}} \rangle_{\alpha},
\end{equation}
where $n_{\text{r}}$ is the isotropic refractive index of the solid, $e$ is the elementary charge, $\epsilon_{0}$ is the permittivity of free space, and $c$ is the speed of light in vacuum. Following mathematical derivations (see Appendix), we obtained the 2D spontaneous recombination coefficient (cm$^2$/s) for a simple non-degenerate two-band model in the limit of low carrier density
\begin{equation}\label{eq:B_2D analyt.}
    B_{\text{2D}} = \frac{}{}
    \frac
    {
        2 q_{\text{e}}^2
    }
    {
        c^3 m_0^2 \epsilon_0 k_{\text{B}}T  
    }
    \,
    \frac
    {
        n_{\text{r}} E_{\text{g}} \langle p^2_{vc,\bm{k}_0} \rangle_{\alpha}
    }
    {
        g_{\bm{k}_0} (m_c + m_v)
    } .
\end{equation}
Here $\bm{k}_0$ denotes the momentum-space location of the valley where direct optical transitions occur in the \ac{BZ}, and $g_{\bm{k}_0}$ is its multiplicity. The final term highlights material-specific parameters, beyond the momentum matrix elements, that influence the rate of radiative recombination. In particular, the presence of multiple $\bm{k}_0$ valleys ($g_{\bm{k}_0} > 1$) as well as heavier effective masses reduce recombination rates by delocalizing carrier density in reciprocal space. Real-world materials often exhibit multiple bands with varying degeneracy near the band edges, as illustrated schematically in Fig.~\ref{fig-band-schematic}, where $\Delta_{c_i}$ and $\Delta_{v_j}$ represent the energy separations of conduction and valence subbands relative to their respective band edges. Equation~\eqref{eq:B_2D analyt.} can be recast in more general form as
\begin{equation}\label{eq:B_2D no-degen. analyt.}
    B_{\text{2D}} = 
    \frac
    {
        2 q_{\text{e}}^2
    }
    {
        c^3 m_0^2 \epsilon_0 k_{\text{B}}T  
    }
    \,
    \frac
    {
        n_{\text{r}} \sum_{i,j} (E_{\text{g}} + \Delta_{c_i} + \Delta_{v_j}) \langle p^2_{v_{j}c_{i},\bm{k}_0} \rangle_{\alpha} \exp[-(\Delta_{c_i} + \Delta_{v_j})/k_{\text{B}}T] \,m_{c_i}m_{v_j}/(m_{c_i} + m_{v_j})
    }
    {
        g_{\bm{k}_0}
        \Bigr[
            \sum_i m_{c_i} \exp(-\Delta_{c_i}/k_{\text{B}}T)
        \Bigr]
        \Bigr[
            \sum_j m_{v_j} \exp(-\Delta_{v_j}/k_{\text{B}}T)
        \Bigr]
    } 
\end{equation}
with the summation indices $i$ and $j$ running over a set of conduction and valence bands, respectively. This expression inherently accounts for band degeneracy.

We evaluate the bimolecular recombination coefficient in \ce{MoS2} in the regime of independent charge carriers (excitonic effects will be discussed later). The average refractive index of $\langle n_{\text{r}} \rangle_{\alpha}=3.6$ was estimated from experimental data~\cite{Ermolaev_NC_12_2021_10.1038/s41467-021-21139-x} based on in-plane and out-of-plane values. The effective masses of $m_c=0.43 m_0$ and $m_v=0.53 m_0$ are sourced from \ac{C2DB}. Electronic structure parameters: $E_{\text{g}}=1.58$~eV, $\Delta_{c_2}=3$~meV, $\Delta_{v_2}=150$~meV, $p_{x/y,c_1 v_1}^2=p_{x/y,c_2 v_2}^2=0.069$~at.u. (other $p_{\alpha,c_i v_j}^2=0$) correspond to the \ac{DFT}-\ac{PBE} level. The multiplicity factor $g_{\bm{k}_0} = 2$ accounts for $K$ and $K'$ valleys in the \ac{BZ}. Using Eq.~\eqref{eq:B_2D no-degen. analyt.} we obtain analytically at room temperature
$$
    B_{\text{2D}}(\text{MoS}_2) = 3.1 \times 10^{-5}~\text{cm}^2/\text{s}.
$$
This result leads to the intrinsic bimolecular radiative lifetime of
$$
    \tau_{\text{bm}}(\text{MoS}_2, n=10^{11}~\text{cm}^{-2}) \sim 300~\text{ns}
$$
for a charge carrier density of $n=10^{11}$~cm$^{-2}$, which serves as a reasonable lower limit, given the constraints expressed in Eq.~\eqref{eq:App:n constraint} and the available experimental data: $10^{10}~\text{cm}^{-2}$~\cite{Wu_2013_10.1021/jz401199x}, $(0.34\!-\!2.1)\times10^{12}~\text{cm}^{-2}$~\cite{Myers_10.1063/5.0213720}, $(0.1\!-\!2.2)\times 10^{12}~\text{cm}^{-2}$~\cite{Qiu2013_10.1038/ncomms3642} and $(4\!-\!12)\times10^{12}~\text{cm}^{-2}$~\cite{Cui2015_10.1038/nnano.2015.70}. 

The bimolecular radiative recombination coefficient can be directly evaluated from \ac{DFT} by approximating the analytical integral in Eq.~\eqref{eq:B(A)} over the full \ac{BZ} as a discrete summation over the $k$-points of the 2D irreducible \ac{BZ}. Using discrete values $f_{\bm{k}}$ of a generic function $f(\bm{k})$, the integral can be expressed as follows~\cite[Eqs.~4.34, 4.44, and 4.35 therein]{Martin2020-fv}
\begin{equation}
    (2\pi)^{-d} \int_{\text{BZ}} f(\bm{k}) ~d\bm{k}
    \approx
    \Omega^{-1}_{\text{cell}}  \sum_{\bm{k}\in\text{IBZ}} w_{\bm{k}} f_{\bm{k}},
\end{equation}
where $w_{\bm{k}} = g_{\bm{k}}/N_{k}$, with $N_{\bm{k}}$ representing the number of $k$-points in the full 2D \ac{BZ}. The weights satisfy the normalization condition $\sum_{\bm{k}\in \text{IBZ}}w_{\bm{k}}=1$. In 2D the symbol $\Omega_\text{cell}$ denotes the lateral area of a 2D primitive cell.

The bimolecular radiative recombination coefficient $B_{\text{2D}}$ and radiative lifetime $\tau_{\text{bm}} =1/(n\,B_{\text{2D}})$ for materials with the highest optical coupling between band edges are listed in Table~\ref{table:best_materials_Bbm}, alongside well-established \acp{TMDC}. Both quantities were normalized by the refractive index to facilitate comparison across systems and reduce uncertainties. The radiative coefficients obtained are comparable to those for the optically active \ce{MoS2} monolayer, indicating their potential for practical optical applications. A correlation is observed between high optical coupling at the band edge and optimal values of experimentally accessible quantities, with additional variability attributed to differences in the electronic structure. Table~\ref{table:best_materials_Bbm} shows that the bimolecular radiative lifetime of independent charge carriers is about two orders of magnitude longer than experimental values for \ce{MoS2} at room temperature: 0.58~ns~\cite{Goodman_10.1103/PhysRevB.96.121404} and 0.85~ns~\cite{Shi_10.1021/nn303973r}.

\begin{table}[]
    \caption{Bimolecular radiative recombination coefficients and intrinsic radiative lifetimes at room temperature for direct band gap \acp{2DM} with the strongest optical coupling at the band edges. The properties are normalized by the refractive index. $B_{\text{2D}}$ and  $\tau_{\text{bm}}$ were computed using \ac{DFT} with the carrier density of $n = 10^{11} \text{cm}^{-2}$ at $T=300$~K. The exciton radiative lifetime was esimated using Eq.~\eqref{eq:exciton lifetime} incorporating the $E_\text{exc}^{(\lambda)}$ and $\langle p_{\text{exc}}^2(\lambda) \rangle_{\alpha}$ characteristics of the brightest exciton states in Table~\ref{table:best_materials}.}\label{table:best_materials_Bbm}
    \begin{ruledtabular}
    \begin{tabular}{lccc}
        Material & $B_\text{2D}/n_{\text{r}}\times10^{5}$ & $\tau_\text{bm}\,n_{\text{r}}$& $ \tau_{\text{exc}}\,n_{\text{r}}$\\
        structure & $(\text{cm}^{2}/\text{s})$ & ($\mu$s)&(ns) \\
        \hline
        GaSeCl   & 2.9  & 0.3& 0.2\\
        BiTeCl   & 1.2  & 0.8& 6\\
        \ce{Bi2Se2Te}  &  0.44 & 2.3& 16\\
        ZrNCl &  1.5 & 0.7& 1\\
        TiNBr  &  0.49 & 2.1& 6\\
        \ce{MoS2}   & 0.41  & 2.5& 3\\
        TiNCl   & 0.42  & 2.4& 6\\
        MoSeS  & 0.45  & 2.2& 4\\
        \ce{MoSe2}  & 0.57  & 1.7& 4\\
    \end{tabular}
    \end{ruledtabular}
\end{table}

\subsection{Excitonic effects}

The results presented above are based on the independent particle approximation within \ac{DFT}, offering an initial framework for studying optical transitions in \acp{2DM}. While higher-level theories are necessary to accurately capture excitonic effects and electronic screening~\cite[chap.20]{Bechstedt_Springer_2016}, \cite[chap.21]{Martin2020-fv}, \cite{Onida_RevModPhys_2002_https://doi.org/10.1103/RevModPhys.74.601}, the high-throughput nature of this study makes solving the \ac{BSE} for all materials impractical. Consequently, we computed and presented exciton characteristics for a selected subset of materials in Table~\ref{table:best_materials}. Low-energy exciton energies are shown side-by-side with corresponding optical couplings obtained at the $GW$-\ac{BSE} level. Inspired by the \ac{BSE} oscillator strength~\cite{Albrecht_PRL_80_1998, Schmidt_PRB_67_2003, Fuchs_PRB_78_2008}, the effective \ac{BSE} momentum matrix element is expressed as a coherent sum of \ac{DFT} momentum matrix elements in the full \ac{BZ} weighted by a \ac{BSE} eigenvector $\mathcal{A}_{v_j c_i,\bm{k}}^{(\lambda)}$ corresponding to the exciton energy $E_{\text{exc}}^{(\lambda)}$
\begin{equation}\label{eq:p2 exciton}
    \begin{split}
        \langle p_{\text{exc}}^2(\lambda) \rangle_{\alpha} 
        & = 
        \frac{1}{3}
        \sum_{\alpha}
        \left|
            \sum_{i,j}
            \sum_{\bm{k}\in\text{FBZ}}
            p_{\alpha,v_j c_i,\bm{k}} \, \mathcal{A}_{v_j c_i,\bm{k}}^{(\lambda)}
        \right|^2 \\
        & =
        \frac{m_0^2}{3\hbar^2}
        \sum_{\alpha}
        \left|
            \sum_{i,j}
            \sum_{\bm{k}\in\text{FBZ}}
            (E_{c_i,\bm{k}} - E_{v_j,\bm{k}})  \bm{r}_{\alpha,v_j c_i,\bm{k}} \, \mathcal{A}_{v_j c_i,\bm{k}}^{(\lambda)}
        \right|^2 \\
        & \approx
        \frac{m_0^2 E_{\text{g}}^2}{3\hbar^2}
        \sum_{\alpha}
        \left|
            \sum_{i,j}
            \sum_{\bm{k}\in\text{FBZ}}
            \bm{r}_{\alpha,v_j c_i,\bm{k}} \, \mathcal{A}_{v_j c_i,\bm{k}}^{(\lambda)}
        \right|^2
        .
    \end{split}
\end{equation}
Here the band gap $E_{\text{g}}$ must be calculated at the same level of theory as dipole matrix elements $\bm{r}_{\alpha,v c,\bm{k}}$ (\ac{DFT}-\ac{PBE} in our case). The approximation in the last line of Eq.~\eqref{eq:p2 exciton} is justified only for the lowest energy excitons and was verified for \ce{MoS2}. The term
$
    \sum_{\alpha}
    \left|
        \sum_{i,j}
        \sum_{\bm{k}\in\text{FBZ}}
        \bm{r}_{\alpha,v_j c_i,\bm{k}} \, \mathcal{A}_{v_j c_i,\bm{k}}^{(\lambda)}
    \right|^2
$
is refered in \ac{VASP} documentation as the \ac{BSE} oscillator strength and stored in \texttt{vasprun.xml} file under the `opticaltransitions' tag in units of~{\AA}$^2$. Across all materials studied at the \ac{BSE} level, bright excitons exhibit stronger optical coupling than in the independent-particle approximation (Table~\ref{table:best_materials}). \ce{Mg2Al2Se5} was included in Table~\ref{table:best_materials} as a benchmark material with inherently weak optical coupling, demonstrating a consistency between the \ac{DFT} and \ac{BSE} levels.

All materials in Table~\ref{table:best_materials} with a band gap at $\Gamma$ exhibit doubly degenerate \ac{CBE} and \ac{VBE} states. Exciton binding energies and momentum matrix elements typically lead to either a 3+1 or 2+2 configuration. The 3+1 case, observed in materials such as GaSeCl and TiNCl, consists of three dark excitons and one bright exciton, analogous to triplet and singlet states. In the 2+2 configuration, both excitons can be bright (e.g., BiTeCl) or a mix of bright and dark (e.g., \ce{Bi2Se2Te}). In many cases, dark and bright excitons are degenerate (e.g., TiNBr, \ce{MoS2}) or nearly so (e.g., ZrNCl), facilitating efficient radiative recombination. However, in some materials, the bright exciton energy exceeds that of the dark exciton by more than $k_{\text{B}} T$ at room temperature. These materials are expected to exhibit characteristics of both direct and indirect semiconductors---similar to hybrid lead halide perovskites---potentially benefiting photovoltaic applications. In hybrid perovskites, this effect arises from Rashba splitting~\cite{Katan_JMCA_3_2015,Azarhoosh_AM_4_2016_10.1063/1.4955028} or dynamic disorder~\cite{Zheng_PRM_2_2018,Munson_C_4_2018}, whereas here it originates from disparities in exciton binding energies. The presence of a dark exciton ground state has been observed in layered lead halide perovskites~\cite{quarti_2024_https://doi.org/10.1002/adom.202202801} and in \acp{TMDC} such as \ce{WS2} and \ce{WSe2}~\cite{Wang_2015_10.1038/ncomms10110,Arora_10.1039/C5NR01536G, Withers_doi:10.1021/acs.nanolett.5b03740}. In the latter case, the dark exciton ground state arises from spin-forbidden optical transitions due to an inverted spin ordering of the two lowest-energy conduction bands in the $K$ valley as compared to \ce{MoS2}~\cite{Echeverry_PhysRevB.93.121107}. Notably, the bulk parent structure of BiTeCl has been previously identified as nanocrystals hosting a bright exciton ground state~\cite{Swift_10.1021/acsnano.4c02905}.

The exciton radiative lifetime can be estimated as the inverse of the Einstein coefficient
\begin{equation}\label{eq:exciton lifetime}
    \tau_{\text{exc}}
    \approx
    \left[
        \frac{
            n_{\text{r}} q_{\text{e}}^2
        }{
            \pi \epsilon_{0} \hbar^{2} m_{0}^{2} c^{3}
        } E_{\text{exc}} \langle p^2_{\text{exc}} \rangle_{\alpha}
    \right]^{-1}.
\end{equation}
Using the values of $E_\text{exc}^{(\lambda)}$ and $\langle p_{\text{exc}}^2(\lambda) \rangle_{\alpha}$ from Table~\ref{table:best_materials} for one of the bright excitons in \ce{MoS2}, we obtain
$\tau_{\text{exc}} \sim 0.8$~ns, in agreement with experimental data~\cite{Goodman_10.1103/PhysRevB.96.121404, Shi_10.1021/nn303973r}. The estimated radiative lifetimes of bright excitons originating from band edges in materials with strong optical coupling are presented in Table~\ref{table:best_materials_Bbm}. As with \ce{MoS2}, these lifetimes fall within the nanosecond range, with an average value of $\tau_{\text{exc}}n_{\text{r}} \sim 5$~ns across materials. These exciton lifetimes are consistently shorter than the bimolecular radiative lifetime of independent charge carriers, suggesting potential advantages for achieving high internal quantum efficiency, provided that non-radiative losses are minimized. These results also indicate promise for the development of optoelectronic devices, such as light-emitting diodes and lasers, where fast radiative recombination is crucial for efficient operation~\cite[chap.~7.1.1]{Cardona_Springer_2016} and \cite[chap.~14.4.2]{Neamen2002}.

\subsection{Validation and limitations}

To benchmark our results for optical transition matrix elements, we selected well-studied materials: bulk GaAs, GaN, and monolayer \ce{MoS2}. The \textit{velocity} matrix elements are chosen for comparison and presented in Table~\ref{tab:p2 comparison} for three different exchange-correlation functionals: \ac{PBE}, \ac{HSE06}~\cite{Heyd_J.Chem._118_18_Phys._10.1063/1.1564060,Krukau_2006_JChemPhys125_22_https://doi.org/10.1063/1.2404663}, and \ac{YSH}~\cite{Tran_Phys.Rev.B_83_23_10.1103/PhysRevB.83.235118}. For local functionals, such as \ac{PBE}, velocity and momentum matrix elements are equivalent to each other ~\cite{Adolph_PhysRevCondMatt_53_15_1996_https://doi.org/10.1103/physrevb.53.9797,Cohen_ChemPhysLett_14_2_1972_https://doi.org/10.1016/0009-2614(72)87178-1,Starace_PRA_3_1971,Rubel_ComputationMPDI_110_2_2022_https://doi.org/10.3390/computation10020022}. In the case of nonlocal functionals, such as \ac{HSE06} and \ac{YSH} hybrid functionals, the velocity operator contains an additional contribution $\langle \phi_{c,\bm{k}}| \mathbbm{i}[\hat{V}_{\text{nl}},\bm{r}] |\phi_{v,\bm{k}}\rangle$ due to the nonlocal nature of the Hartree-Fock potential. This additional contribution to the velocity matrix elements, inherent to hybrid functionals, assures their better agree with experimental values, which are obtained from the energy parameter $E_P$ derived from empirical band structure measurements. While \ac{DFT}-\ac{PBE} underestimates the optical transition strength by approximately 35\% compared to both experimental data and hybrid functionals, this discrepancy is systematic. Consequently, trends in the relative efficiency of optical transitions at the band edges, as predicted by \ac{DFT}-\ac{PBE}, are expected to remain valid.

\begin{table}
    \footnotesize
    \caption{Velocity matrix elements $v^2_{x,vc} = \sum_{i,j} v^2_{x,v_j c_i}$ (at.u.) computed at two different levels of theory and compared with experiment. Degeneracy of bands is indicated in brackets.}
    \label{tab:p2 comparison}
    \begin{ruledtabular}
        \begin{tabular}{lcccc}
            Material & Transitions & \ac{PBE} & Hybrid & Experiment \\
            & & (this work) & (\ac{HSE06} and \ac{YSH})  \\
            \hline
            GaAs (bulk) & $\Gamma_{\text{so}}^{(\times 2)},\Gamma_{lh,hh}^{(\times 4)}\to \Gamma_{c}^{(\times 2)}$ & 0.62 & 0.80~\cite{Rubel_ComputationMPDI_110_2_2022_https://doi.org/10.3390/computation10020022} & 0.83$-$1.1~\cite{Aspnes_PhysRevB_14_https://doi-org.libaccess.lib.mcmaster.ca/10.1103/PhysRevB.14.5331,Pfeffer_PhysRevB_52_19_1996_https://doi.org/10.1103/physrevb.53.12813,Hermann_PhysRev_15_2_1977_https://doi.org/10.1103/physrevb.15.823,Cardona_PhysRevB_38_3_1988_https://doi.org/10.1103/physrevb.38.1806,Vurgaftman_J.Appl.Phys_89_11_2001_https://doi.org/10.1063/1.1368156}\\
            GaN (bulk, wurtzite) & $\Gamma_{\text{so}}^{(\times 2)}, \Gamma_{lh}^{(\times 2)}, \Gamma_{hh}^{(\times 2)} \to \Gamma_{c}^{(\times 2)}$ & 0.36 & 0.49~\cite{De_Carvalho_PhysRevB_84_19_2011_https://doi.org/10.1103/physrevb.84.195105,Rubel_ComputationMPDI_110_2_2022_https://doi.org/10.3390/computation10020022} & 0.51$-$0.69~\cite{Shokhovets_AppPhysLett_86_16_2005_https://doi.org/10.1063/1.1906313,Shokhovets_PhysRevBCondensedMattMatPhys_78_3_2008_https://doi.org/10.1103/physrevb.78.079902,Vurgaftman_J.Appl.Phys_89_11_2001_https://doi.org/10.1063/1.1368156} \\
            \ce{MoS2} (monolayer) & $K_{v_1,v_2} \to K_{c_1,c_2}$ & 0.14 & 0.21~\cite{Rubel_ComputationMPDI_110_2_2022_https://doi.org/10.3390/computation10020022} & --- \\
        \end{tabular}
    \end{ruledtabular}
\end{table}

The energy parameter $E_P$ is suitable for benchmarking purposes, however, caution is advised when utilizing it for predicting the efficiency of radiative recombination~\cite{Im_APL_70_1997_10.1063/1.118293}. This parameter accounts for coupling between the conduction band and valence states, including the split-off band ($\Gamma_{\text{so}}\to \Gamma_{c}$ in GaAs and GaN). However, the split-off band has a negligible contribution to radiative optical transitions at room temperature due to its low thermal occupation. The high excess energy ($\Delta_{\text{so}} > k_{\text{B}}T$) results in a reduced population of carriers in the split-off band at low carrier densities, leading to an approximately 30\% reduction in the effective momentum matrix element $\langle p_{vc}^2 \rangle_{\alpha,300~\text{K}}$ relative to the values listed in Table~\ref{tab:p2 comparison}.

\section{Conclusions}

We conducted a comprehensive benchmarking of optical transition matrix elements in direct band gap \acp{2DM}, including well-studied \acp{TMDC}. Using a temperature-dependent summation approach, we ranked materials based on the matrix elements of the lowest-energy transitions that dominate spontaneous emission, accounting for thermal population factors. Our analysis identified transition-metal nitrogen halides (ZrNCl, TiNBr, TiNCl) and bismuth chalcohalides (BiTeCl) with optical coupling at the band edges comparable to or exceeding that of \ce{MoS2}. Some of these materials exhibit direct band gaps in the visible range, making them promising for optoelectronic applications. However, despite strong in-plane dipole transitions, most \acp{2DM} underperform compared to bulk semiconductors due to the absence of out-of-plane optical components.

To better understand interband transitions, we introduced the orbital overlap tensor as a quantitative metric. Our analysis revealed a significant fraction of $d$-$d$ transitions in TMDCs, which are nominally forbidden by atomic selection rules yet exhibit strong optical coupling that increases with ionicity of the bond. Conventional explanations based on spin-orbit coupling and crystal field splitting proved insufficient. Instead, we established a correlation between anomalous \acp{BEC} and interband optical coupling, linking charge redistribution dynamics to optical transition strength. Real-space visualization of optical dipoles confirmed their localization at transition-metal sites in \ce{MoS2}, ruling out conventional orbital hybridization as the primary factor in optical coupling.

We derived an analytical expression for the radiative recombination coefficient in \acp{2DM}, explicitly incorporating multi-valley and multi-band effects. However, radiative lifetimes computed within the independent-particle approximation were two orders of magnitude longer than experimental values for \ce{MoS2}. This discrepancy was resolved by including excitonic effects at the $GW$-\ac{BSE} level for a selected subset of materials. Notably, we observed a correlation between the effective \ac{BSE} momentum matrix element and \ac{DFT}-derived properties, suggesting that preliminary \ac{DFT}-based screening can significantly reduce computational costs. However, even materials with direct band gaps and strong optical coupling at the \ac{DFT} level can exhibit dark excitons as the lowest-energy states, making them quasi-direct band gap semiconductors—a phenomenon analogous to hybrid lead halide perovskites but driven by excitonic effects rather than Rashba splitting or dynamic disorder. This insight is critical for engineering excitonic recombination dynamics in next-generation optoelectronic materials.

\begin{acknowledgments}
Authors are thankful to Peter Blaha (TU Vienna) for suggesting to plot the orbital overlap. A.~F.~G.-B. and O.~R. acknowledge funding provided by the Natural Sciences and Engineering Research Council of Canada under the Discovery Grant Programs RGPIN-2020-04788 (application id 5017093). K.~S. acknowledges the funding provided by the Mitacs Globalink program. Computing resources were provided by the Digital Research Alliance of Canada. A.~C.~G.~C. acknowledges the grant entitled ``Búsqueda y estudio de nuevos compuestos antiperovskitas laminares con respuesta termoeléctrica mejorada para su uso en nuevas energías limpias" supported by Vicerrectoría de Investigaciones y Extensión, VIE--UIS.

\textbf{Author contributions}: A.~F.~G.-B. refined the optical matrix element calculations and performed convergence tests, extracted and processed data, performed \ac{DFT} calculations for bimolecular recombination coefficients, co-developed the orbital overlap tensor, proposed the connection between opposite Bader and Born charge behavior with momentum matrix elements, and drafted the manuscript. K.~S. carried out preliminary optical matrix element calculations and initial data processing. A.~C.~G.~C. contributed to discussions and provided editorial revisions. O.~R. conceptualized the project, derived analytical expressions for bimolecular recombination coefficients, proposed the real-space visualization of optical activity, introduced the orbital overlap tensor, performed $GW$-\ac{BSE} calculations, and extensively restructured and rewrote most of the manuscript.

\textbf{Data availability}:
The \ac{VASP} input files required to reproduce the calculations and the scripts used for the computation of the momentum matrix elements and excitonic properties are available in the Zenodo repository~\cite{Zenodo_10.5281/zenodo.13773214}.

\textbf{Supporting material}:
Supplemental material is available with the list of pseudopotentials used, convergence tests, and Bader charges of select materials. The following references are cited within the supplemental material section ~\cite{Min_JChemphys_10.1063/1.3553716,HENKELMAN_https://doi.org/10.1016/j.commatsci.2005.04.010,Tang_2009_10.1088/0953-8984/21/8/084204,Sanville_https://doi.org/10.1002/jcc.20575, Haastrup_2DMATER_5_2018_https://doi.org/10.1088/2053-1583/aacfc1, Gjerding_2DMaterials_8_4_2021_https://dx.doi.org/10.1088/2053-1583/ac1059}.

\end{acknowledgments}

% \appendix*
\begin{appendix}
\section{Analytical derivation of $B_\text{2D}$}\label{Sec:Appendix}

Here, we present the analytical derivation of the bimolecular recombination coefficient for a \ac{2DM}. For simplicity, we assume there is only one valence band and one conduction band (both non-degenerate) with isotropic band dispersion. The extension of this result to multiple bands with additional splittings in energy is straightforward. We will also neglect any possible variation of the momentum matrix element $\langle p^2_{vc,\bm{k}} \rangle_{\alpha}$ as a function of $\bm{k}$ in the vicinity of the band extrema $\bm{k}_0$. By combining Eqs.~\eqref{eq:B(A)} and \eqref{eq:A}, we obtain our starting point
\begin{equation}\label{eq:App:B general}
    B_\text{2D} = 
    \frac
    {
        n_{\text{r}} q_{\text{e}}^2
    }{
        \pi \epsilon_{0} \hbar^{2} m_{0}^{2} c^{3}
    }
    \,
    \langle p^2_{vc,\bm{k}_0} \rangle_{\alpha}
    (2\pi)^{-2} n^{-2} g_{\bm{k}_0}
    \int_{0}^{\infty} f_{c,k}(1-f_{v,k})(E_{c,k}-E_{v,k})
    \, 2\pi k~dk ,
\end{equation}
where we used $d\bm{k} \to 2\pi k~dk$ in a 2D \ac{BZ} with isotropic band dispersion. The integral is evaluated in the vicinity of the band extremum $\bm{k}_0$. The factor $g_{\bm{k}_0}$ represents multiplicity of $\bm{k}_0$ in the whole \ac{BZ}. The band dispersion is expressed in terms of effective masses $m_c$ and $m_v$ as follows
\begin{equation}
    E_{c,k} = E_{\text{g}} + \frac{\hbar^2 k^2}{2m_c}
\end{equation}
and
\begin{equation}
    E_{v,k} = -\frac{\hbar^2 k^2}{2m_v} .
\end{equation}
The occupancy of states in the conduction band, in the limit of a low areal density of optical excitations $n$ (the limit will be rigorously defined at the end of this section), can be approximated as
\begin{equation}
    f_{c,k} = \frac{1}{1+\exp[(E_{c,k}-\mu_c)/k_{\text{B}}T]}
    \approx e^{(\mu_c-E_{c,k})/k_{\text{B}}T} ,
\end{equation}
and similarly for the valence band as
\begin{equation}
    1-f_{v,k} = 1 - \frac{1}{1+\exp[(E_{v,k}-\mu_v)/k_{\text{B}}T]}
    \approx e^{(E_{v,k}-\mu_v)/k_{\text{B}}T} .
\end{equation}
The occupancy product in Eq.~\eqref{eq:App:B general} becomes
\begin{equation}\label{eq:App:fc(1-fv)}
    f_{c,k}(1-f_{v,k}) = e^{(\mu_c-\mu_v)/k_{\text{B}}T} \, e^{(E_{v,k}-E_{c,k})/k_{\text{B}}T} .
\end{equation}
The quasi Fermi levels $\mu_c$ and $\mu_v$ can be related to the areal density of optical excitations in the conduction band as
\begin{equation}\label{eq:App:n_c}
    n = (2\pi)^{-2} g_{\bm{k}_0}  \int_{0}^{\infty} f_{c,k} \, 2\pi k~dk \approx
    \frac{g_{\bm{k}_0}m_c k_{\text{B}}T}{2\pi \hbar^2} \,
    e^{\mu_c/k_{\text{B}}T} e^{-E_{\text{g}}/k_{\text{B}}T} ,
\end{equation}
and the same density in the valence band is given by
\begin{equation}\label{eq:App:n_v}
    n = (2\pi)^{-2} g_{\bm{k}_0}  \int_{0}^{\infty} (1-f_{v,k}) \, 2\pi k~dk \approx
    \frac{g_{\bm{k}_0} m_v k_{\text{B}}T}{2\pi \hbar^2} \,
    e^{-\mu_v/k_{\text{B}}T} .
\end{equation}
Combining Eqs.~\eqref{eq:App:n_c} and \eqref{eq:App:n_v} results in the exponential term expressed as
\begin{equation}\label{eq:App:exp term}
    e^{(\mu_c-\mu_v)/k_{\text{B}}T}
    = \frac{(2\pi)^2 \hbar^4 n^2}{g_{\bm{k}_0}^2 m_c m_v (k_{\text{B}}T)^2} \,
    e^{E_{\text{g}}/k_{\text{B}}T} .
\end{equation}
Note that $e^{(\mu_c-\mu_v)/k_{\text{B}}T} \propto n^2$, which leads to the elimination of the carrier concentration from the final expression for $B_{\text{2D}}$ in the limit of a low density of optical excitations. With Eqs.~\eqref{eq:App:fc(1-fv)} and \eqref{eq:App:exp term}, the integral in Eq.~\eqref{eq:App:B general} can be evaluated analytically
\begin{equation}
    \begin{split}
        & (2\pi)^{-2} g_{\bm{k}_0} \int_{0}^{\infty} f_{c,k}(1-f_{v,k})(E_{c,k}-E_{v,k}) \, 2\pi k~dk \\
        & \approx (2\pi)^{-1} g_{\bm{k}_0} \, e^{(\mu_c-\mu_v)/k_{\text{B}}T} \int_{0}^{\infty} e^{(E_{v,k}-E_{c,k})/k_{\text{B}}T}(E_{c,k}-E_{v,k}) \, k~dk \\
        & = \frac{2\pi\hbar^2 n^2(E_{\text{g}} + k_{\text{B}}T)}{g_{\bm{k}_0} k_{\text{B}}T(m_c+m_v)} .
    \end{split}
\end{equation}

After substituting this result back into Eq.~\eqref{eq:App:B general} and noting that $E_{\text{g}} \gg k_{\text{B}}T$, we obtain a final expression for the bimolecular recombination coefficient in SI units, in the limit of low density of optical excitations, with material-dependent terms grouped at the end
\begin{equation}\label{eq:App:B_2D analyt.}
    B_{\text{2D}} = 
    \frac
    {
        2 q_{\text{e}}^2
    }
    {
        c^3 m_0^2 \epsilon_0 k_{\text{B}}T  
    }
    \,
    \frac
    {
        n_{\text{r}} E_{\text{g}} \langle p^2_{vc, \bm{k}_0} \rangle_{\alpha}
    }
    {
        g_{\bm{k}_0} (m_c + m_v)
    } .
\end{equation}
The low-density approximation in Eqs.~\eqref{eq:App:n_c} and \eqref{eq:App:n_v} has less than 5\% error as long as the areal density of charge carriers satisfies
\begin{equation}\label{eq:App:n constraint}
    n \lesssim 0.017 \, g_{\bm{k}_0} k_{\text{B}}T \, \hbar^{-2} \min(m_v, m_c) .
\end{equation}
This expression yields the upper limit of $5 \times 10^{11}$~cm$^{-2}$ for the low carrier density at room temperature, with $g_{\bm{k}_0}=2$ and a mass of $0.43 m_0$, which are parameters relevant to \ce{MoS2}. The expression derived for the 2D bimolecular recombination coefficient [Eq.~\eqref{eq:App:B_2D analyt.}] is in agreement with that given by~\citet[Appendix B.5]{Blood2015_Oxforf_2015} in the context of quantum wells for double-degenerate bands.

 For comparison, the spontaneous recombination coefficient (cm$^3$/s) in bulk materials, in the limit of low carrier density, is
\begin{equation}\label{eq:App:B_3D analyt.}
    B_{\text{3D}} = 
    \frac
    {
        2 \sqrt{2 \pi} q_{\text{e}}^2 \hbar
    }
    {
        c^3 m_0^2 \epsilon_0 (k_{\text{B}}T)^{3/2} 
    }
    \,
    \frac
    {
        n_{\text{r}} E_{\text{g}} \langle p^2_{vc,\bm{k}_0} \rangle_{\alpha}
    }
    {
        g_{\bm{k}_0} (m_c + m_v)^{3/2}
    } ,
\end{equation}
which is consistent with the expressions quoted earlier by~\citet{Im_APL_70_1997_10.1063/1.118293} and by~\citet[Eq.~(4.5.23)]{Landsberg_Recombination}, with the exception of differences in numerical factor of two that accounts for double-degenerate bands. Differences between $B_{\text{3D}}$ and $B_{\text{2D}}$ are related to the effect of dimensionality on the joint density of states, weighted by the Fermi-Dirac occupation statistics.
\end{appendix}

\bibliography{bibliography}

\clearpage
%\makeatletter

\renewcommand\thesection{S\arabic{section}}
\renewcommand\thefigure{S\arabic{figure}}
\renewcommand\thetable{S\arabic{table}}
\renewcommand\theequation{S\arabic{equation}}
\renewcommand\thepage{S\arabic{page}}
%\renewcommand\thetitle{Supplementary information for publication}
%\makeatother
\setcounter{section}{0}
\setcounter{figure}{0}
\setcounter{table}{0}
\setcounter{equation}{0}
\setcounter{page}{1}
%\appendix

\title{Supplementary information: Efficiency of band edge optical transitions of 2D monolayer materials: A high-throughput computational study}
\maketitle

\section*{Pseudopotential valency and cutoff energy information}

\begin{longtable}{@{\extracolsep{5pt}}lcc}
\caption{Pseudopotential \texttt{POTCAR} file names according to \ac{VASP} convention, along with their respective cutoff energies and number of valence electrons considered.}
\label{table:SI_Potcars_1}\\
\hline
\hline 
Atom & Cutoff energy (eV) & Valency      \\
\hline
H         & 250                 & 1       \\
Li\_sv    & 499                 & 3       \\
C         & 400                 & 4       \\
N         & 400                 & 5       \\
O         & 400                 & 6       \\
F         & 400                 & 7       \\
Na\_pv    & 260                 & 7       \\
Mg\_pv    & 404                 & 8       \\
Al        & 240                 & 3       \\
Si        & 245                 & 4       \\
P         & 255                 & 5       \\
S         & 259                 & 6       \\
Cl        & 262                 & 7       \\
K\_sv     & 259                 & 9       \\
Ca\_sv    & 267                 & 10      \\
Sc\_sv    & 223                 & 11      \\
Ti\_sv    & 275                 & 12      \\
V\_sv     & 264                 & 13      \\
Cr\_pv    & 266                 & 12      \\
Mn\_pv    & 270                 & 13      \\
Fe        & 268                 & 8       \\
Co        & 268                 & 9       \\
Ni        & 270                 & 10      \\
Cu\_pv    & 369                 & 17      \\
Zn        & 277                 & 12      \\
Ga\_d     & 283                 & 13      \\
Ge\_d     & 310                 & 14      \\
As        & 209                 & 5       \\
Se        & 212                 & 6       \\
Br        & 216                 & 7       \\
Rb\_sv    & 220                 & 9       \\
Sr\_sv    & 229                 & 10      \\
Zr\_sv    & 230                 & 12      \\
Nb\_sv    & 293                 & 13      \\
Mo\_sv    & 243                 & 14      \\
Ru\_pv    & 240                 & 14       \\
Rh\_pv    & 247                 & 15      \\
Pd        & 251                 & 10      \\ 
Ag        & 250                 & 11      \\
Cd        & 274                 & 12      \\
In\_d     & 239                 & 13      \\
Sn\_d     & 241                 & 14      \\
Sb        & 172                 & 5       \\
Te        & 175                 & 6       \\
I         & 176                 & 7       \\
Cs\_sv    & 220                 & 9       \\
Ba\_sv    & 187                 & 10      \\
Hf\_pv    & 220                 & 10      \\
Ta\_pv    & 224                 & 11      \\
W\_sv     & 223                 & 14      \\
Re        & 226                 & 7       \\
Os        & 228                 & 8       \\
Ir        & 211                 & 9       \\
Pt        & 230                 & 10      \\
Au        & 230                 & 11      \\
Hg        & 233                 & 12      \\
Tl\_d     & 237                 & 13      \\
Pb\_d     & 238                 & 14      \\
Bi\_d     & 243                 & 15      \\
%\end{tabular}
%\end{ruledtabular}
\hline
\hline
\end{longtable}
%\clearpage

\section*{Convergence tests and bader charges}

\begin{figure}[hbt!]
    \centering
    \includegraphics{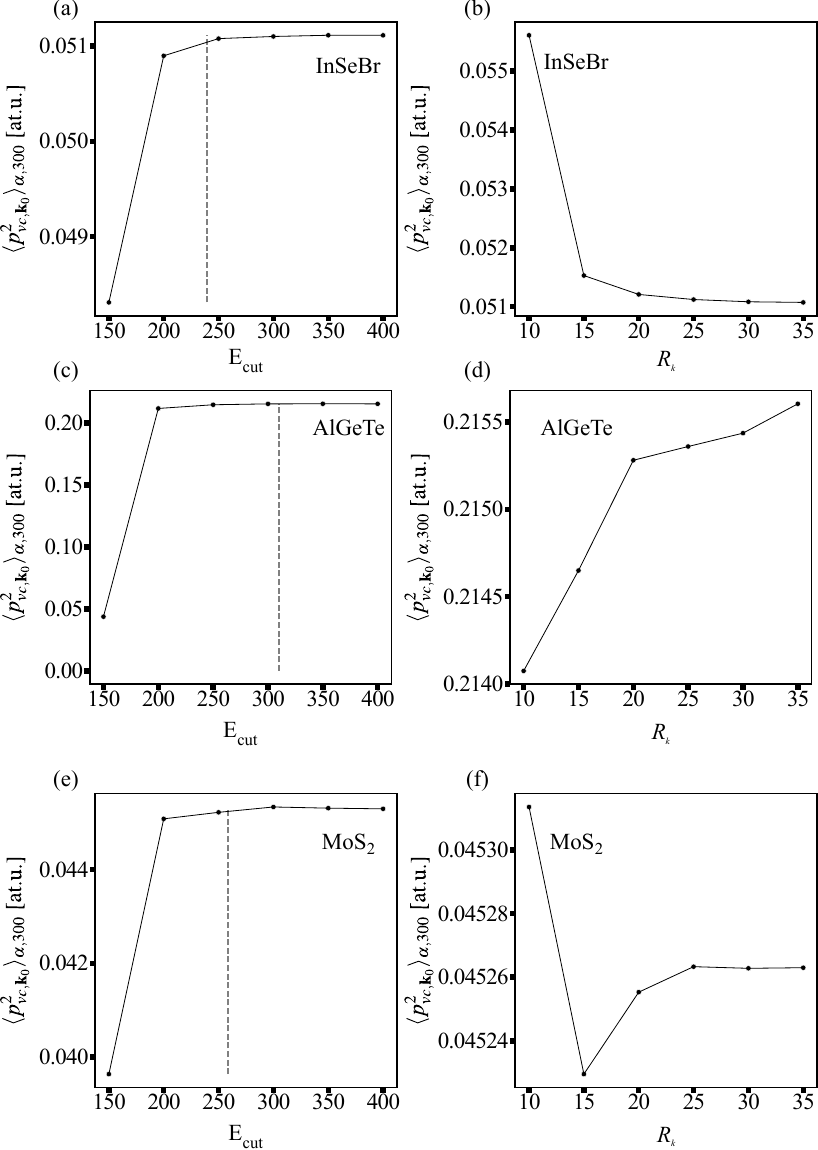}
    \caption{Momentum matrix elements are plotted as a function of the computational parameters used for the \ac{DFT} calculations. Dashed lines represent the cutoff energy values employed in the production run. $R_k=20$ was used in the production run except for $GW$-\ac{BSE} calculations.} 
    \label{sup:fig-convergence}
\end{figure}

Figure~\ref{sup:fig-convergence} presents convergence studies of the in-plane momentum matrix elements plotted against the cuttoff energy and $R_{k}$ computational parameters for three representative materials from the \acp{2DM} ensemble. A $k$-mesh of $15\times15\times1$ was employed in  convergence studies of the matrix elements with respect to the cutoff energy. The dashed lines indicate the values selected for each case in the high throughput calculation. When examining convergence of the momentum matrix elements with respect to the density of $k$-points, the cutoff energy and the precision mode used in these calculations were consistent with the details provided in Sec.~\ref{sec:Computational and theoretical details} of the main text. Additionally, calculations utilizing \ac{VASP}'s normal precision mode (\texttt{PREC=normal}) were performed for comparative purposes. This selection results in a less dense Fourier grid. The differences between the squared momentum matrix elements obtained using the normal mode and the accurate mode were found to be immaterial.

In Table~\ref{table:SI:Bader}, we present the Bader charges computed for \ce{CrMo3Te8}, \ce{Mo2W2Te8}, \ce{Mo3WTe8} and \ce{MoW3Te8} \acp{2DM}, which are not reported in \ac{C2DB}. For comparison, we also present the \ac{BEC}s, which exhibit an opposite sign to the Bader charges in all instances, exhibiting the anomalous behavior. The Bader charges were computed employing the \texttt{bader} software~\cite{Min_JChemphys_10.1063/1.3553716,HENKELMAN_https://doi.org/10.1016/j.commatsci.2005.04.010,Tang_2009_10.1088/0953-8984/21/8/084204,Sanville_https://doi.org/10.1002/jcc.20575}. We employed the same computational parameters outlined in Sec.~\ref{sec:Computational and theoretical details} of the main text. A significant parameter for the calculation of Bader charges is the `fine' Fourier grid (\texttt{PREC=accurate}), which is utilized for the representation of localized augmentation charges. To validate the results, the Bader charges were computed on denser grids, as detailed in Table~\ref{table:SI:Bader_conver}. The charges are converged within approximately 5\%, which is sufficient for our purpose.  

\clearpage

\begin{table}[]
\caption{Computed Bader charges (e), and the average of the diagonal in-plane components of the \ac{BEC} (e) obtained from \ac{C2DB}~\cite{Haastrup_2DMATER_5_2018_https://doi.org/10.1088/2053-1583/aacfc1, Gjerding_2DMaterials_8_4_2021_https://dx.doi.org/10.1088/2053-1583/ac1059}. }
\label{table:SI:Bader}
\begin{tabular}{cc}
\hline
\hline
                                         & \ce{CrMo3Te8}                                                                              \\ \hline
$Z^{A}_{Bader}$                          &  0.63, 0.52, 0.55, 0.55, -0.28, -0.28, -0.28, -0.28, -0.27, -0.29, -0.27, -0.29        \\
$\langle Z^{A}_{xx}, Z^{A}_{yy} \rangle$  & -3.83,  -3.43,  -3.38,  -3.38,  1.46,  1.46,  1.83,  1.83,  1.83,  1.84,  1.89,  1.90 \\ \hline
\multicolumn{1}{l}{}                     & \ce{Mo2W2Te8}                                                                            \\ \hline
$Z^{A}_{Bader}$                           &  0.54, 0.54, 0.54, 0.54, -0.26, -0.28, -0.26, -0.28, -0.27, -0.28, -0.27, -0.28        \\
$\langle Z^{A}_{xx}, Z^{A}_{yy} \rangle$  & -3.38,  -3.37,  -2.52,  -2.52,  1.41,  1.41,  1.41,  1.41,  1.54,  1.54,  1.54,  1.54 \\ \hline
\multicolumn{1}{l}{}                     & \ce{Mo3WTe8}                                                                               \\ \hline
$Z^{A}_{Bader}$                           &  0.54, 0.55, 0.54, 0.47, -0.26, -0.28, -0.25, -0.28, -0.25, -0.26, -0.25, -0.26        \\
$\langle Z^{A}_{xx}, Z^{A}_{yy} \rangle$  & -3.31,  -3.31,  -3.26,  -2.54,  1.52,  1.52,  1.52,  1.52,  1.54,  1.54,  1.63,  1.64 \\ \hline
\multicolumn{1}{l}{}                     & \ce{MoW3Te8}                                                                               \\ \hline
$Z^{A}_{Bader}$                           &  0.55, 0.53, 0.54, 0.54, -0.27, -0.27, -0.27, -0.27, -0.26, -0.27, -0.26, -0.27        \\
$\langle Z^{A}_{xx}, Z^{A}_{yy} \rangle$  & -3.39,  -2.62,  -2.56,  -2.54,  1.32,  1.32,  1.40,  1.41,  1.42,  1.42,  1.42,  1.42 \\ 
\hline
\hline
\end{tabular}
\end{table}

\begin{table}[]
\caption{Computed Bader charges (e) for \ce{CrMo3Te8} employing different Fourier grid densities. $\text{NG(X, Y, Z)F}_{in}$ refers to the initial grid constructed from the \texttt{ENCUT} value.} 
\label{table:SI:Bader_conver}
\begin{tabular}{cc}
\hline
\hline
FFT grid              & $Z^{A}_{Bader}$                                                               \\ \hline
 $\text{[NG(X, Y, Z)F]}_{in}$          & 0.63, 0.52, 0.55, 0.55, -0.28, -0.28, -0.28, -0.28, -0.27, -0.29, -0.27, -0.29 \\ \hline
2$\times$$\text{[NG(X, Y, Z)F]}_{in}$ & 0.66, 0.58, 0.57, 0.57, -0.29, -0.29, -0.29, -0.29, -0.29, -0.30, -0.29, -0.30 \\ \hline
3$\times$$\text{[NG(X, Y, Z)F]}_{in}$ & 0.68, 0.59, 0.59, 0.59, -0.30, -0.31, -0.31, -0.31, -0.30, -0.31, -0.30, -0.31 \\
\hline
\hline
\end{tabular}
\end{table}

\end{document}